\documentclass[prd,preprint,a4paper,showkeys,showpacs]{revtex4-1} 

\usepackage{bm}
\usepackage{amsmath}
\usepackage{amssymb}
\usepackage{graphicx}
\usepackage{color}
\usepackage{tikz}
\usepackage{color}

\def\p{\partial}

\def\b{\beta}
\def\d{\delta}

\def\e{\epsilon}
\def\ve{\varepsilon}

\def\l{\lambda}
\def\L{\Lambda}

\def\f{\phi}

\def\ra{\rightarrow}

\def\Red{}

\begin{document}

\title{
Effective sextic field theory for tricritical-critical crossover}
\author{Jos\'e Gaite}
\affiliation{
Applied Physics Dept.,
ETSIAE,
Universidad Polit\'ecnica de Madrid,\\ E-28040 Madrid, Spain}

\date{May 10, 2026}

\begin{abstract}
Effective field theories provide a suitable framework for both particle physics and statistical 
physics. We delve deeper into the study of the effective three-dimensional scalar 
field theory for its application to statistical physics, 
especially considering the role of the sextic coupling in the tricritical-to-critical crossover. 
The three-loop renormalization of the mass and the two coupling constants that we perform 
allows us to obtain, for the first time, the complete renormalization group flow of the couplings 
in that order. We analyze what universality means in this problem and how we can recover 
non-universal terms from the renormalization group beta functions. 
The crossover is realized by the convergence of the  renormalization group flow towards the 
line connecting the tricritical and critical fixed points. 
\end{abstract}


\maketitle

\section{Introduction}
\label{intro}

Renormalization theory is a venerable topic in quantum field theory and statistical physics 
\cite{I-Z,Parisi,ZJ}. Its basic assumption is that infinite quantities appearing in perturbative 
calculations belong to an unobservable domain, while observable renormalized quantities are always 
finite. In the practical application of renormalization in field theory, the use of dimensional 
regularization and the epsilon expansion has been standard for many years \cite{ZJ}. This method 
allows for a straightforward derivation of the renormalization group (RG).
However, a different approach is gaining considerable popularity, namely, the set of methods 
that constitute {\em effective field theory} \cite{Cao-S,Weinberg,Burgess}.

In effective field theory, unlike in dimensional regularization, a wavenumber or small distance 
cutoff is introduced (let us assume it is of the former type and call it $\L$). Furthermore, 
this cutoff is not removed but held fixed. One is interested in the physics at smaller 
scales. In this approach, one can dispense with counterterms, the BPHZ procedure, etc, which are 
standard in traditional renormalization theory \cite{I-Z}. 
On the other hand, the methods of effective field theory can be 
non-perturbative and are then equivalent to the Wilson renormalization group approach 
\cite{Wil-Kog}, in which renormalization is understood as the effect of the progressive removal of 
short-distance degrees of freedom, implemented by means of a running cutoff. 
Effective field theory accounts for
the effect of those degrees of freedom through a small number of {\em relevant} 
coupling constants. The theory of a scalar field with only relevant couplings is considered 
{\em universal}, since it is determined by symmetry and space dimension only. 

In effective field theory, physical observables are functions of the cutoff $\L$. 
However, the physics on scales far below $\L$ must be mildly dependent on it. 
In fact, physical observables are functions of the ratio 
of physical mass to cutoff $m/\L$, with definite limit when $m/\L\ra 0$. 
Of course, the existence of the limit $\L\ra\infty$ is what is called renormalizibility 
in traditional renormalization theory. 
In effective field theory, instead of renormalizibility, universality is considered, defined as 
independence of $\L$. 
However, corrections to universality can be considered, and even why non-universal 
quantities should not have physical consequences. Universality lies in the behavior under the renormalization group: relevant couplings tend to definite limits, which 
can be independently fixed, whereas irrelevant couplings tend to become 
functions of the relevant couplings. 
In reality, most of the non-universal quantities represent short-range degrees of freedom that 
belong to an unobservable domain and are therefore irrelevant. Consequently, in the
opposite direction of the RG, the irrelevant couplings appear, the number of couplings increases, 
and in fact, it becomes arbitrarily large on the cutoff scale.

The $(\lambda\phi^4)_3$ theory (the three-dimensional quartic scalar field theory) has long been 
studied, especially as the Ising model universality class \cite{Parisi,ZJ}. This universality class
has only one relevant parameter, because $\l/m$ tends to a definite value as $m/\L$ decreases,  
making $\l$ irrelevant. On the other hand, 
the complete $(\lambda\phi^4+g\phi^6)_3$ theory (a.k.a.\ the sextic field theory), which 
contains all the power-counting renormalizable couplings, has additional interest. 
Indeed, this theory has been studied for various reasons, especially for
the description of tricritical phenomena 
\cite{LL,Migdal,R-Wegner,Stephen-M,Wegner-R,R-Wegner_1,Stephen,Goro,Carvalho,Pfeu-T,Soko,Law-Sar,Ginzburg,Hager,Zin-Cod,AKT}, 
but also for other reasons \cite{McK,HT,Huish,Shrock,Kharuk}.
In the sextic field theory, as $m/\L$ decreases, we have two possibilities. The first one is that 
both $\l$ and $m$ are relevant parameters, such that they remain independent as $m$ decreases, 
while $g$ becomes a function of them. This corresponds to the 
tricritical universality class. 
The second possibility is that $\l$ also becomes a function of $m$, which 
corresponds to ordinary critical bahavior (in which the sextic coupling is superfluous 
from the outset).

Therefore, the sextic field theory can describe both critical and tricritical behavior; moreover, it 
can describe the {\em crossover} from one to the other. However, the tools of perturbative field 
theory needed for this purpose are not fully developed.  Most authors have employed dimensional 
regularization and the epsilon expansion in $d=3-\ve$ dimensions 
\cite{McK,HT,Huish,Zin-Cod,Shrock,Hager,AKT}, which are adequate for describing some aspects of 
tricritical phenomena in $d=3$ but cannot describe the crossover to critical behavior (we shall 
explain why). This is not a limitation unique to dimensional regularization: Kharuk employs cutoff 
regularization \cite{Kharuk} but omits several indispensable terms (actually emphasizing consistency 
with the results of dimensional regularization).

We have recently shown that the effective sextic field theory constitutes a suitable framework for 
deducing important properties of three-dimensional scalar field theory and thus for describing both critical and tricritical behavior \cite{I}. In particular, we have obtained, for the first time, 
the complete RG beta functions of the $(\lambda\phi^4+g\phi^6)_3$ theory up to the two-loop order, 
thereby extending and adding rigor to previous calculations \cite{Goro,Soko}.  Our approach, in 
Ref.~\onlinecite{I}, focuses on scheme dependence, and we find it remarkable that the obtained 
beta functions are scheme independent (universal). Therefore, it is possible to recover universal 
properties of three-dimensional scalar field theory from the beta functions. Furthermore, we have 
shown that a small scheme dependence appears in effective field theory when the condition $m \ll \L$ 
fails \cite{I}.

Here we develop the methods of Ref.~\onlinecite{I}, extending them to the three-loop order
and performing a detailed analysis of the RG flow. 
To do this effectively, we attempt to simplify the renormalization scheme 
as much as possible. This simplification allows us to find a common basis for 
dimensional regularization and cutoff schemes (a cutoof is required in effective 
field theory). Thus, we can make contact with the calculations of perturbative sextic field theory 
that employ dimensional regularization and the epsilon expansion in $d=3-\ve$ dimensions 
\cite{McK,HT,Huish,Zin-Cod,Shrock,Hager,AKT}. Furthermore, we can reveal the crucial 
ingredients for the description of the tricritical-to-critical crossover.

The contents is structured in six main sections and a concluding section. 
After this last section, there are three appendices, the first one containing the calculation of 
some necessary integrals, the second exposing the long formulas obtained in the first 
renormalization step, and the third summarizing the calculation of the beta-functions. 
Section \ref{phi4} is also introductory and quotes some known results of three-dimensional 
$\lambda\phi^4$ theory. Section \ref{3loop} introduces the three-loop effective potential and 
sets up the renormalization procedure. 
The results are given in the next section, Sec.~\ref{ren+} (with additional comments).
In Sec.~\ref{RG}, we show the beta functions and plot the RG flow. We also make some 
comments on the connection with related beta functions. 
An analysis of the RG by means of the theory of dynamical systems is carried out in 
Sec.~\ref{dynamical}. Some further properties of the RG, especially, those that 
are relevant to the tricritical-critical crossover, are studied in 
Sec.~\ref{genRG}. 

\section{Renormalization of the $(\lambda\phi^4)_3$ theory}
\label{phi4}

Scalar field theory in three-dimensional space, with one or several fields, 
has many applications in statistical and condensed matter physics 
and is widely studied \cite{Parisi,ZJ}.
The discovery of the Wilson-Fisher fixed point and the $\e$ expansion provided a 
suitable method of calculating universal quantities 
\cite{Wil-Kog}. However, the calculation in fixed dimension ($d=3$) has some 
advantages \cite{Parisi,ZJ}. For example, 
the $(\lambda\phi^4)_3$ theory is super-renormalizable, 
the {\em superficial} UV divergences of Feynman graphs are limited to the two-loop order 
and they only require a mass renormalization \cite[ch 9]{ZJ}. 
Therefore, using the renormalized mass, the perturbation theory can be expressed in terms of 
finite integrals, which can be calculated numerically. 
These properties allowed the early five-loop computation of Baker et al \cite{Baker}
(which included the Pad\'e-Borel resummation of the power series).  

The well-known six-loop $\b$-function of $(\lambda\phi^4)_3$ theory 
\cite[\S 29.2]{ZJ} has been recently extended to the 
seven-loop order \cite{SS}. We just quote, as a reference, the four-loop result of 
the coupling constant renormalization and the beta function
($u=\l/m$ is the dimensionless coupling constant):
\begin{align}
\frac{\l_0}{\l} = 1+\frac{9 u}{2 \pi }+\frac{575 u^2}{36 \pi ^2}+1.83421
   u^3+1.91785 u^4+O\left(u^5\right),
\label{l0lf4}
\\
\b(u) = -u+\frac{9 u^2}{2 \pi }-\frac{77 u^3}{9 \pi ^2}+1.03177
   u^4-1.58506 u^5+O\left(u^6\right).
\label{betaf4}
\end{align}
(The latter matching the beta fuction in Ref.~\onlinecite[\S 29.2]{ZJ}, 
after taking into account the different normalization of the coupling constant.) 

The RG fixed point, $\b(u^*)=0$, is found at $u^*=0.985$, 
by means of the six-loop beta function \cite[\S 29.4]{ZJ}. This value gives a rough idea of 
where naive perturbation theory must fail. In fact, we find that the perturbative results 
for $u$, e.g. Eq.~(\ref{l0lf4}), are reliable even for $u\gtrsim 0.5$, although the 
expansion of $g(u)$ is not \cite{I} (the function $g(u)$ is studied in Sec.~\ref{separatrices}).

Here we are interested in the 3-loop effective potential and we will use Rajantie's 
analytic calculation of the vacuum Feynman graph integrals up to the 3-loop order \cite{Raj}. 
In addition, we will use the analytic expression of the field renormalization factor 
to the 3-loop order \cite{Kudlis} (not calculated by Rajantie). 
These calculations allow us to express the first four terms of expansions~(\ref{l0lf4})
and (\ref{betaf4}) in explicit analytic form. We will only need two additional 
Feynman graph integrals for the sextic theory.

In $(\lambda\phi^4)_3$ theory, the sextic term $\f^6$ is a {\em marginal} composite field that hardly 
alters the results for the quartic coupling \cite{Parisi,ZJ}. In fact, this marginal coupling is 
actually irrelevant for critical behavior but it has some influence on tricritical behavior, 
which is described by the {\em classical} Landau theory with small corrections, 
as we shall see.

The ordinary critical behavior is ruled by the non-trivial RG fixed point of 
$(\lambda\phi^4)_3$ theory, which is the well understood Wilson-Fisher fixed point found in 
the $\e$ expansion. It is also found in fixed dimension, as a zero of beta function 
Eq.~(\ref{betaf4}), for $u^*\simeq 1$ 
(thus demanding a series summation procedure) \cite{Parisi,ZJ}. In contrast, 
tricritical behavior is ruled by the trivial RG fixed point $u=0$, but requires 
that we consider it in the complete $(\lambda\phi^4+g\phi^6)_3$ theory \cite[\S 24.7]{ZJ}, 
\cite[Sec.\ III]{Law-Sar}.

\section{Three-loop perturbative $(\lambda\phi^4+g\phi^6)_3$ theory} 
\label{3loop}

The effective potential of three-loop order can be calculated with 
the background-field method, as done for two-loop order in Ref.~\onlinecite{I}. 
We need eight three-loop vacuum Feynman graphs in addition to the 3 Feynman graphs already 
present in the two-loop order \cite{I}. The full set of graphs is displayed below.

\begin{center}
	\begin{tikzpicture}
	\draw [thick] (0,0) circle (0.5);
	\draw [thick] (2.5, 0.4) circle (0.4) ;
	\draw [fill=black] (2.5,0) circle (2pt);
	\draw [thick] (2.5,-0.4) circle (0.4) ;
	\draw [thick] (5,0) circle (0.5);	
	\draw [fill=black] (4.5,0) circle (2pt);
	\draw [fill=black] (5.5,0) circle (2pt);
	\draw [thick] (4.5,0) -- (5.5,0);
	\end{tikzpicture}\qquad
	\raisebox{8pt}{
		\begin{tikzpicture}
	\draw [thick] (0,0) circle (0.5);
	\draw [thick] (-0.5,0) arc (240:300:1);
	\draw [thick] (0.5,0) arc (60:120:1);
	\draw [fill=black] (0.5,0) circle (2pt);
	\draw [fill=black] (-0.5,0) circle (2pt);
	\end{tikzpicture}\quad
		\raisebox{3pt}{\begin{tikzpicture}
	\draw [thick] (-0.40, 0) circle (0.40) ;
	\draw [thick] (0.40,0) circle (0.40) ;
	\draw [thick] (1.2,0) circle (0.4) ;
	\draw [fill=black] (0,0) circle (2pt);
	\draw [fill=black] (.8,0) circle (2pt);
	\end{tikzpicture}}}\quad
	\raisebox{6pt}{	\begin{tikzpicture}
	\draw [thick] (-0.5, 0) circle (0.5) ;
	\draw [thick] (0.5,0.5) -- (0.5,-0.5);
	\draw [thick] (0.5,0) circle (0.5) ;
	\draw [fill=black] (0,0) circle (2pt);
	\draw [fill=black] (0.5,0.5) circle (2pt);
	\draw [fill=black] (0.5,-0.5) circle (2pt);
	\end{tikzpicture}\quad
		\begin{tikzpicture}
	\draw [thick] (0,0) circle (0.5);
	\draw [thick] (-0.5,0) arc (90:36:1);
	\draw [thick] (-0.5,0) arc (270:324:1);
	\draw [fill=black] (-.5,0) circle (2pt);
	\draw [fill=black] (0.3,0.38) circle (2pt);
	\draw [fill=black] (0.3,-0.38) circle (2pt);
	\end{tikzpicture}}\\
		\begin{tikzpicture}
	\draw [thick] (0,0) circle (0.5);
	\draw [thick] (0,0.17) arc (270:295:1);
	\draw [thick] (0,0.17) arc (270:245:1);
	\draw [thick] (0,-0.17) arc (90:65:1);
	\draw [thick] (0,-0.17) arc (90:115:1);
	\draw [fill=black] (0.41,0.26) circle (2pt);
	\draw [fill=black] (0.41,-0.26) circle (2pt);
	\draw [fill=black] (-0.41,0.26) circle (2pt);
	\draw [fill=black] (-0.41,-0.26) circle (2pt);
	\end{tikzpicture}\qquad
	\begin{tikzpicture}
	\draw [thick] (0,0) circle (0.5);
	\draw [thick] (0,0) -- (0,0.5);
	\draw [thick] (0,0) -- (0.43,-0.25);
	\draw [thick] (0,0) -- (-0.43,-0.25);
	\draw [fill=black] (0,0) circle (2pt);
	\draw [fill=black] (0,0.5) circle (2pt);
	\draw [fill=black] (0.43,-0.25) circle (2pt);
	\draw [fill=black] (-0.43,-0.25) circle (2pt);
	\end{tikzpicture}\qquad
	\begin{tikzpicture}
	\draw [thick] (-0.5, 0) circle (0.5) ;
	\draw [thick] (0,0) -- (1,0);
	\draw [thick] (0.5,0) circle (0.5) ;
	\draw [fill=black] (0,0) circle (2pt);
	\draw [fill=black] (1,0) circle (2pt);
	\end{tikzpicture}\qquad	
	\raisebox{5pt}{\begin{tikzpicture}
	\draw [thick] (0,0.5) arc (120:240:0.305);
	\draw [thick] (0,0.5) arc (60:-60:0.305);
	\draw [thick] (0,0) arc (60:180:0.305);
	\draw [thick] (0,0) arc (0:-120:0.305);
	\draw [thick] (0,0) arc (120:0:0.305);
	\draw [thick] (0,0) arc (180:300:0.305);
	\draw [fill=black] (0,0) circle (2pt);
	\end{tikzpicture}}
\end{center}

The combinatorial factors are not difficult to obtain and we arrive at:
\begin{align}
U_\mathrm{eff}(\f) &= U(\f)
+ \frac{h}{2}\, I_1
+ h^2 \left(\frac{3\, U^{(4)}(\f)}{4!}\, I_2^2
-\frac{3 \,U^{(3)}(\f)^2 }{(3!)^2}\,I_3 \right)
\nonumber\\
%
   &+ h^3
   \left\{-\left(\frac{U^{(4)}(\f)}{4!}\right)^2 4\cdot 3\,(I_4 + 3\, I_5)
   + \frac{U^{(4)}(\f)}{4!} \left(\frac{U^{(3)}(\f)}{3!}\right)^2
   {4}\cdot{27}\,(I_6 + I_7) \right.  \nonumber\\ 
   &- \left.
   \left(\frac{U^{(3)}(\f)}{3!}\right)^4 {27}
   \left(3\,I_8 + 2\,I_9\right)  
   -\frac{U^{(5)}(\f)}{5!} \frac{U^{(3)}(\f)}{3!} {60}\, I_{10} 
   +\frac{U^{(6)}(\f)}{6!} \,15\,  I_{11} 
   \right\} + O\!\left(h^4\right).
\label{Vef}
\end{align}
The last two terms are present for the sextic potential 
but vanish for the quartic potential (because $U^{(5)}=U^{(6)}=0$). 
We have eleven integrals, $I_1,\ldots,I_{11}$, corresponding to the Feynman graphs drawn above. 
Of course, $I_{10}=I_2\,I_3$ and $I_{11}=I_2^3$. The integrals are to be calculated with a massive 
correlation function such that $M^2 = U''(\f)$.

In addition, we have to consider the contribution of some of the above Feynman graphs to 
field renormalization. We need precisely the graphs with just two fields attached 
to different vertices (two background fields attached to the same vertex or to a line in some graph 
only produce a mass contribution). Besides, we can only attach one background field to vertices with 
three or five internal lines. 
Therefore, the only contributing graphs are the third, the sixth, 
the seventh, and the tenth ones. The sixth graph only differs from the third one in that 
it contains a one-loop mass contribution. 

Naturally, the field renormalization factor $Z$ corresponds to space-varying background fields or, 
alternatively, to fields dependent on the wavenumber (in Fourier space). In the latter case, the 
integrals to be calculated have a similar appearance to those of the effective potential, but are 
more difficult, since they include a non-zero external wavenumber at the corresponding vertex pair.

\subsection{Calculation of integrals for the effective action}

Most of the integrals are divergent and need regularization. In the spirit of effecive field theory, 
we use a cutoff $\L$ on either real or wavenumber space 
(the usual convention is that $\L$ represents a wavenumber cutoff, while a 
real space cutoff is often denoted by $1/\L$). 
The two-loop integrals have been calculated before, e.g., in Ref.~\onlinecite{I}, employing various 
cutoff schemes. 
Of course, the simplest regulating procedure is dimensional regularization, in which 
terms corresponding to powers of $\L$ do not appear, remaining just dimension poles, 
corresponding to $\log\L$ terms. 
Rajantie \cite{Raj} calculated all the 3-loop integrals for the quartic potential in 
dimensional regularization in the MS-scheme with $\ve=3-d$, obtaining analytic expressions 
in terms of elementary and dilogarithm functions. There are only two divergent integrals, namely, 
$I_3$ and $I_4$, which correspond to the third and fourth Feynman graphs above. 

Naturally, the sextic potential produces one more divergent integral 
in dimensional regularization, $I_{10}=I_2\,I_3$, but it does not demand any further calculation. 
The divergent part of 3-loop integrals in dimensional regularization is not especially difficult 
to calculate and it had been calculated earlier by other people, e.g., Huish and Toms \cite{HT}. 
The finite integrals can be computed numerically; indeed, numerical integrals 
for the quartic theory up to six-loop order had been published by Nickel el al 
\cite{Nickel-etal}. However, we find it useful to employ 
the analytic expressions of the most difficult 
3-loop integrals which have been provided by Rajantie \cite{Raj}, for the effective potential, and 
by Kudlis-Pikelner \cite{Kudlis}, for the field renormalization. 
Unfortunately, these expressions are rather long and involve transcendental numbers and special 
functions. We provide the complete expressions in appendix \ref{integrals}. We do not need 
to calculate additional integrals for the sextic theory, since the last two Feynman graphs mentioned 
above do not require them. 

As noticed by Rajantie \cite{Raj}, most integrals are simpler in real space, because 
the (bare) real-space correlation function is a simple function in three dimensions. 
Naturally, we can also take advantage of the simple form of this function in cutoff regularization, 
which is the appropriate type of regularization in effective field theories. 
In Ref.~\onlinecite{I}, several calculations of two-loop integrals with various 
cutoff schemes were made in real space. 
paying close attention to positive powers of the cutoff $\L$ 
(these are the ones most affected 
when the cutoff scheme changes). 
Nevertheless, the two-loop RG beta functions turned out to be scheme independent \cite{I}. 
Furthermore, it was shown that the integration of the RG flow between an initial large mass $m_0$ 
and a final physical mass $m$ should recover the cutoff scheme 
dependence, provided that $m_0$ is identified with $\L$. In this sense, it is unnecessary to 
consider complex cutoff schemes, if we only want to establish the 
characteristics of the RG flow. Moreover, we can argue that positive powers of $\L$, though 
making large contributions if $m \ll \L$, are not really meaningful, and we only need 
$\log\L$ terms, that is, terms equivalent to pole terms in dimensional regularization.

The argument for discarding positive powers of $\L$ because they make no contribution to 
the beta functions was already presented by Sokolov \cite{Soko} (albeit in a slightly different 
way, since he did not employ the background-field method). We should also mention that by
relaxing the conditions on cutoff schemes that we imposed before, in Ref.~\onlinecite{I}, one can 
find regulators that cause the positive powers to naturally cancel out in the renormalization 
formulas. These regulators are called enhanced regulators \cite{Padi}. 

Our own argument relies on the notion of {\em normal-ordered vertex},  
considered by 
Refs.~\onlinecite{I-Z} and \onlinecite[A10.2]{ZJ}. Powers of $\L$ arise in the 
self-contractions of the vertex $\f^4$ in the super-renormalizable quartic theory, but they 
can be eliminated a priori by replacing the vertex $\f^4$ by its normal-ordered vertex, denoted 
$:\!\phi^4\!:$ \cite[A10.2]{ZJ}. This procedure can be extended to the sextic theory, and thus we can 
define:
\begin{align}
U(\phi) &= \frac{m_c^2 \phi^2}{2} + \l_c :\!\phi^4\!: + g_c :\!\phi^6\!: \,= 
\frac{m_c^2 \phi^2}{2} + \l_c \left(\phi^4 - 6 \,h\langle \phi^2\rangle \phi^2 + 
3\, h^2\langle \phi^2\rangle^2\right) 
\nonumber\\
&\phantom{==} + g_c \left(\phi^6 - 15\, h\langle \phi^2\rangle \phi^4 + 45\, h^2\langle \phi^2\rangle^2 \phi^2
- 15\, h^3\langle \phi^2\rangle^3 \right) \nonumber\\
&= 
\left(\frac{m_c^2}{2} - 6 \,h\l_c \langle \phi^2\rangle 
+ 45\, h^2 g_c \langle \phi^2\rangle^2\right) \phi^2 
+ \left(\l_c - 15\, h g_c \langle \phi^2\rangle\right) \phi^4 + g_c \,\phi^6
\nonumber\\
&\phantom{==} + 
3 \, h^2\l_c \langle \phi^2\rangle^2 - 15 \, h^3 g_c \langle \phi^2\rangle^3.
\end{align}
(The subscript c is chosen to mean that this redefinition removes ``cactus'' Feynman graphs). 

Therefore, the normal-order divergences can be absorbed through a partial renormalization:
\begin{align}
m^2 &= m_c^2 - 12 \, h\l_c \langle \phi^2\rangle_\mathrm{div} + 
90\, h^2\, g_c \langle \phi^2\rangle^2_\mathrm{div}\,, 
\nonumber\\
\l &= \l_c - 15\, h\, g_c \langle \phi^2\rangle_\mathrm{div}.
\label{mlc}
\end{align}
We are assuming a sort of {\em minimal subtraction}, in which we just subtract 
positive powers of $\L$, given by $\langle \phi^2\rangle_\mathrm{div}$ and $\langle \phi^2\rangle^2_\mathrm{div}$ (a more general scheme, with an arbitrary mass parameter $\mu$, 
is considered by Zinn-Justin \cite[A10.2]{ZJ}). 
Let us notice that $\langle \phi^2\rangle^2_\mathrm{div} \neq 
\left(\langle \phi^2\rangle_\mathrm{div}\right)^2$. Indeed, while 
$\langle \phi^2\rangle_\mathrm{div}$ 
is just the term of $I_2$ proportional to $\L$, the divergent part of $I_2^2$ that constitutes 
$\langle \phi^2\rangle^2_\mathrm{div}$ 
contains both $\L^2$ and $\L$ terms (which depend on the cutoff scheme \cite[app.\ A]{I}). 
Furthermore, we have $I_{11}=I_2^3$, which also contains a $\L^3$ term. However, this 
term is already subtracted by Eq.~(\ref{mlc}), as well as any ``cactus'' divergence at higher 
loop order. 

As regards normal-order divergences, we only need the (minimally-subtracted) finite part of $I_2$, 
that is,
$$
\left.I_2\right|_\mathrm{MS} = -\frac{M}{4\pi}.
$$
Powers of $I_2$ are to be replaced by powers of this finite part. Let us notice that 
this finite part is cutoff-scheme independent \cite[app.\ A]{I}. Besides, it  
cannot be absorbed by some non-minimal subtraction, because it is 
field dependent, as given by $M^2 = U''(\f)$. 
 
After dealing with normal-order divergences, we still have a positive power of $\L$ in $I_{4}$ 
(calculated in appendix \ref{integrals}). It is multiplied by $U^{(4)}(\f)^2$ and yields 
$O\left(h^3\right)$ contributions that only depend on the bare coupling constants and 
$\L$. Therefore, it will not contribute to the beta functions, which are calculated as derivatives 
with respect to $m$, while holding bare coupling constants and $\L$ constant 
(Sect.~\ref{RGbetas} and appendix \ref{betas}).

The upshot is that we only have to consider   
$\log\L$ terms, which correspond to $\ve$-poles in dimensional regularization. Such terms arise 
in the integrals $I_3$ and $I_4$ from the two sunset-type Feynman graphs (appendix \ref{integrals}). 
These terms are the only ones that contribute to the beta functions in dimensional regularization, 
because they are calculated as derivatives with respect to $\mu$ 
\cite{McK,HT,Huish,Zin-Cod,Shrock,Hager,AKT}. In our case, finite terms that depend on $m$ 
also contribute. These terms are crucial to describe the critical to tricritical crossover. 

%

Taking all the above points into account and employing the expressions in appendix \ref{integrals}, 
we can write:
\begin{align}
U_\mathrm{eff}(\f) &= U(\f)
-\frac{h}{12 \pi } \,U''(\f)^{3/2}
+\frac{h^2}{384 \pi ^2} \left\{3 U^{(4)}(\f) U''(\f)+2 U^{(3)}(\f)^2 
   \left(A+\log\sqrt{U''(\f)}\right)\right\}  \nonumber\\
   &+ \frac{h^3}{64 \pi^3} 
   \left\{-\frac{U^{(4)}(\phi)^2}{12} \sqrt{U''(\phi )} \left(B+\log
   \sqrt{U''(\phi )}\right)
   +\frac{{k_3}\, U^{(3)}(\phi )^2 U^{(4)}(\phi )}{216
   \sqrt{U''(\phi )}}
   \right. \nonumber\\
   &\left.\phantom{==}   
   +\frac{{k_4}\, U^{(3)}(\phi )^4}{1296\,U''(\phi)^{3/2}}
   -\frac{U^{(5)}(\phi ) U^{(3)}(\phi )}{12} \sqrt{U''(\phi)} \left(A+\log\sqrt{U''(\f)}\right)
   \right. \nonumber\\
   &\left.\phantom{==}
   -\frac{U^{(6)}(\phi ) \,U''(\phi )^{3/2}}{48}\right\}+O\left(h^4\right)
\label{Veff}
\end{align}
This expression, when restricted to $O\left(h^2\right)$, can be compared with the one obtained 
before \cite[Eq.\ 8]{I}. Naturally, terms with powers of the cutoff
do not appear anymore. Furthermore, in comparison with Ref.~\onlinecite{I}, 
we have reversed the sign of the constant $A$ and 
absorbed $-\log\L$ in this constant, to simplify the notation.
The rationale is that both $A$ and $\log\L$ can be bunched together, as non-universal constants 
(the value of $A$ is undefined before fixing the cutoff scheme). In fact, 
being $\L$ a fixed scale in effective field theory, it can be taken 
as the reference scale and therefore set to $\L=1$, as in Ref.~\onlinecite{I}  
(so that dimensional quantities become numbers and $\log\L=0$).

In the 3-loop order, 
the combination of $I_4$ and $I_5$ gives rise to a new constant in Eq.~(\ref{Veff}): 
$$B=A-5/8+\log(4/3).$$ 
For the calculation of the finite integrals 
$I_7$, $I_8$, and $I_9$, we use Rajantie's results, 
in terms of some special functions \cite{Raj}. We borrow  
the notation of Guida and Zinn-Justin \cite{Guida-ZJ}, with the same constants $k_3$ and $k_4$ 
defined by them (see appendix \ref{integrals}). 

To set up the complete 3-loop renormalization of the sextic theory, we still need 
the field renormalization factor $Z$, given by the second derivative of the effective action 
with respect to a space-varying background field, named $\Gamma^{(2)}(\f)$. 
In Fourier space:
\begin{equation}
\Gamma^{(2)}(0) = Z^{-1} \left(m^2 + p^2\right) + O(p^4).
\label{Zdef}
\end{equation}
As analyzed above, the graphs contributing to $\Gamma^{(2)}$ are the third, the sixth, 
the seventh, and the tenth ones, with integrals dependent on the external wavenumber $p$ 
(the sixth graph integral is deduced from the third graph integral).

The full function $\Gamma^{(2)}(p,\f)$ for the quartic theory has been calculated by 
Rajantie \cite{Raj}, but only up to the two-loop order (it is already a fairly complex function, 
which is non-local in real space). 	 
Terms pertinent to the sextic theory have also been calculated, either in dimensional regularization 
with $\overline{\mathrm{MS}}$ \cite{McK,HT,Hager,Zin-Cod,Shrock,AKT} or in regularization with 
spatial cutoff \cite{Kharuk}. These references consider higher-order terms that contribute to $Z$. 
However, their calculations discard finite terms that are relevant to us. Indeed, they only show 
an $O\left(h^4\right)$ term, which would make $Z=1$ in our order. 
The 3-loop calculation of $Z$ for the quartic theory 
is set forth by Kudlis and Pikelner \cite{Kudlis}. They calculate 
the third and seventh Feynman graph integrals, while the sixth is automatically included 
in their scheme (note that the third Feynman graph has two loops and is standard). 
For the sextic theory, we further need the tenth Feynman graph integral, 
which is derived directly from the third integral (see appendix \ref{I4Z}).

\section{Renormalization}
\label{ren+}

Let us introduce in Eq.~(\ref{Veff}) for $U_\mathrm{eff}$ the sextic potential 
\begin{equation}
U(\f) = \frac{m_0^2}{2}\,\f^2 + \l_0\,\f^4 + g_0\,\f^6,
\label{pot}
\end{equation}
with bare mass $m_0$, and bare couplings $\l_0$ and $g_0$. 
We define, as usual, their renormalized values:
\begin{align}
Z^{-1} m^2= U_\mathrm{eff}''(0) &= m_0^2 + O\left(h\right),
\label{mtm0}\\
Z^{-2} \lambda = \frac{U_\mathrm{eff}^{(4)}(0)}{4!} &= \lambda_0 + O\left(h\right),
\label{ll0}\\
Z^{-3} g = \frac{U_\mathrm{eff}^{(6)}(0)}{6!} &= g_0 + O\left(h\right).
\label{gg0}
\end{align}
The right-hand side expressions are easy to derive but are lengthy and are therefore 
included in appendix \ref{fren}. 
They agree with the expressions in Ref.~\onlinecite{I}, up to the two-loop order and 
by suppressing terms with powers of the cutoff. 
Given that we now consider the field renormalization, unlike in Ref.~\onlinecite{I}, 
we have the renormalization factor $Z$, which is also a function of the mass and the coupling 
constants. 
For this reason, it is convenient to use the 
{\em intermediate scheme} mass \cite{Kudlis}: 
\begin{equation}
\widetilde{m}= Z^{-1/2}m. 
\label{mtm}
\end{equation}
In terms of it, we can proceed with Eq.~(\ref{mtm0}) as in Ref.~\onlinecite{I} (in fact, 
$Z \simeq 1$ so $\widetilde{m} \simeq m$). First, we 
invert this equation (up to the 3-loop order) to obtain $m_0$ as a function of 
$\widetilde{m}$ and bare couplings (without knowing $Z$). Then, substituting for $m_0$ in 
Eqs.~(\ref{ll0}) and (\ref{gg0}) will eliminate $m_0$.  
Let us recall that $m_0$ has a dubious physical meaning, given that we may have that 
$m_0^2<0$ when $m\ra 0$ \cite{I}. 

Therefore: 
\begin{align}
Z^{-2}\l &= 
\lambda _0 - \frac{h}{4 \pi}\left(15 g_0 \widetilde{m}+\frac{18 \lambda _0^2}{\widetilde{m}}\right) +
\frac{h^2}{16 \pi^2}\left(60 g_0 \lambda _0 (8 A+8 \log \widetilde{m}+9)+\frac{396 \lambda
   _0^3}{\widetilde{m}^2}\right)
\nonumber\\
&\phantom{==} +{}
\frac{h^3}{64 \pi^3}\left\{
-\frac{20 g_0 \lambda _0^2}{\widetilde{m}} \left(864 \log
   \widetilde{m}+864 A-80 {k_3}+531+468 \log
   \left(\frac{4}{3}\right)\right) -
\right.
\nonumber\\
&
\left.   
450 g_0^2
   \widetilde{m} \left(40 \log \widetilde{m} +40 A-15+24
   \log \left(\frac{4}{3}\right)\right)
   -\frac{8 \lambda _0^4}{\widetilde{m}^3}
   \left(48 {k_3}-32 {k_4}+621-108 \log
   \left(\frac{4}{3}\right)\right)
\right\} \nonumber\\
&\phantom{==} +{} O\!\left(h^4\right),
\label{ltl0}
\end{align}
\begin{align}
Z^{-3}g = g_0 &+ \frac{h}{4 \pi}\left(\frac{36 \lambda _0^3}{\widetilde{m}^3}-\frac{90 g_0 \lambda _0}{\widetilde{m}}\right) +
\frac{h^2}{16 \pi^2}\left(
150 g_0^2 \left(8 \log \widetilde{m}+8
   A+9\right)+\frac{4140 g_0 \lambda
   _0^2}{\widetilde{m}^2}-\frac{2376 \lambda _0^4}{\widetilde{m}^4}\right)
\nonumber\\
&+ 
\frac{h^3}{64 \pi^3}\left\{
\frac{160 g_0 \lambda _0^3}{\widetilde{m}^3} \left(324 \log
   \widetilde{m}+324 A-66 {k_3}+32
   {k_4}-378+189 \log
   \left(\frac{4}{3}\right)\right)
\right.
\nonumber\\
&
\left. \phantom{===}
   {}- \frac{100 g_0^2 \lambda _0}{\widetilde{m}}
    \left(1512 \log \widetilde{m} + 1512 A-112
   {k_3}+981+864 \log
   \left(\frac{4}{3}\right)\right)
\right.
\nonumber\\
&
\left.   \phantom{===}
   {}+ \frac{288 \lambda _0^5}{\widetilde{m}^5} \left(12 {k_3}-16 {k_4}+195-18 \log
   \left(\frac{4}{3}\right)\right)
   \right\}  + O\!\left(h^4\right),
\label{gtg0}
\end{align}
We have replaced $B=A-5/8+\log(4/3)$, leaving only $A$ as scheme-dependent constant 
(which includes the $\log\L$ term). 
Let us notice the following: if we restrict ourselves to the super-renormalizable 
quartic theory, with $g_0=0$, then 
the substitution of $m_0$ by $\widetilde{m}$ has indeed removed the 
scheme dependence in this order,  
because $Z$ is scheme independent (see below).

\subsection{Field renormalization}

Equations (\ref{ltl0}) and (\ref{gtg0}), together with Eq.~(\ref{mtm}), constitute the full 
renormalization equations, once we know $Z$. 
Following the procedure explained in the preceding section and gathering the results, we obtain:
\begin{equation}
Z^{-1} = 
1 + 
h^2\,\frac{\lambda_0^2}{9{\pi}^2\widetilde{m}^2} - h^3\,k_5\,\frac{\lambda_0^3}{2{\pi}^3\widetilde{m}^3} 
\Red{{}-h^3\,\frac{5\l_0 g_0}{6\,{\pi}^3\widetilde{m}} + O\left(h^4\,\right)}. 
\label{Zi}
\end{equation}
The new constant $k_5$ is taken from Ref.~\onlinecite{Kudlis}:
$$
k_5 = 72 \,\text{Li}_2\left(\frac{1}{3}\right) - 3\pi ^2-8+18 \log ^2 3 +32 \log 3-64 \log 2 = 1.2777943312499929.
$$

Now we have the function $Z^{-1}(\l_0/\widetilde{m},g_0)$ but we intend to express everything 
in terms of the physical mass $m$, employing Eq.~(\ref{mtm}).
Therefore, we have to solve for $Z^{-1}$ in the equation:
$$
Z^{-1}[\l_0/(Z^{-1/2}m),g_0] = 1 + 
h^2\,\frac{\lambda_0^2}{9{\pi}^2Z^{-1}m^2} - h^3\,k_5\,\frac{\lambda_0^3}{2{\pi}^3Z^{-3/2}m^3} 
\Red{{}-h^3\,\frac{5\l g_0}{6\,{\pi}^3 Z^{-1/2}m} + O\left(h^4\,\right)}.
$$
It can be transformed into a polynomial equation for $Z^{1/2}$. However, it is simpler and 
consistent with the loop expansion to define $\tilde{\l}_0= \lambda_0/{m}$ and assume
$$
Z^{-1} = 1 + z_1 h + z_2 h^2 + z_3 h^3 + O\left(h^4\,\right),
$$
in terms of functions $z_i(\tilde{\l}_0,g_0)$. We can thus solve for these functions order by order. 
Of course, $z_1=0$, and we can write:
\begin{align*}
1 + z_2 \,h^2 + z_3\, h^3 + O\left(h^4\,\right) 
&= 1 + 
h^2\,\frac{\tilde{\l}_0^2}{9{\pi}^2} \left(1 + z_2 h^2\right)^{-1} 
- h^3\,k_5\,\frac{\tilde{\l}_0^3}{2{\pi}^3} -h^3\,\frac{\Red{5}\tilde{\l}_0 g_0}{6\,{\pi}^3} 
+ O\left(h^4\,\right) \\
&= 1 + 
h^2\,\frac{\tilde{\l}_0^2}{9{\pi}^2} 
- h^3\,k_5\,\frac{\tilde{\l}_0^3}{2{\pi}^3} -h^3\,\frac{\Red{5}\tilde{\l}_0 g_0}{6\,{\pi}^3} 
+ O\left(h^4\,\right).
\end{align*}
Therefore, up to this order, we can simply replace $\widetilde{m}$ with $m$ in Eq.~(\ref{Zi}). 
In addition, Eq.~(\ref{mtm}) writes:
\begin{equation}
\widetilde{m}= Z^{-1/2}m = 
m+\frac{h^2 \lambda _0^2}{18 \pi ^2 m}- \Red{\frac{h^3}{12 \pi^3} 
\left(5 g_0 \lambda _0 +\frac{3 {k_5} \lambda _0^3}{m^2}\right)} + O\left(h^4\right).
\label{mtm+}
\end{equation}
As remarked earlier, $Z \simeq 1$, because the deviation starts in the second order and has small 
numerical coefficients.

Equations (\ref{ltl0}), (\ref{gtg0}), (\ref{Zi}), and (\ref{mtm+}) embody the renormalization 
process, which must be understood from the perspective of effective field theories.

\subsection{Renormalizing the effective field theory}

In effective field theory, the cutoff $\L$ is held fixed and divergences 
occurring in the limit $\L \ra \infty$ are not considered. The nature of the relationship between 
bare mass and coupling constants and their renormalized counterparts differs from what 
is traditional in perturbative field theory, where renormalization absorbs all short-distance 
divergences and allows the $\L \ra \infty$ limit to be taken \cite[\S 12.4]{Weinberg}. 
In effective field theory, 
the effect of field fluctuations on scales between $\L$ and $m$ ($m < \L$) is 
absorbed by renormalization. 
Naturally, the appearance of any effect related to the scale $\L$ is undesirable, 
because it reveals {\em non-universal} features that should not appear,
especially when $m \ll \L$. 
Ultimately, any dependence on $\L$ must disappear, as in traditional field theory. Therefore, 
there is no essential mathematical difference, 
but the interpretation is different and richer \cite[\S 12.4]{Weinberg}.

The standard renormalization method relies on obtaining 
\Red{counterterms, usually through dimensional regularization,} 
and yields expressions for bare coupling constants in terms of renormalized mass and coupling 
constants. It is understood that bare coupling constants tend to infinity as 
the regularization is removed.  
In contrast, we will combine equations (\ref{ltl0}), (\ref{gtg0}), (\ref{Zi}), 
and (\ref{mtm+}), 
to obtain instead the renormalized coupling constants in terms of the renormalized mass and 
bare coupling constants, without ever having to consider infinities (we can set $\L=1$). 
Of course, these equations can be inverted to obtain the bare coupling constants. Nevertheless, 
the renormalized coupling constants as functions of the renormalized mass and 
bare coupling constants provide a straightforward derivation of the RG beta functions 
\cite{I}. 

In addition, some expressions in terms of bare coupling constants are useful to 
compare with known results. For example, we can compare with known results of the quartic theory  
by just setting $g_0=0$. 
Indeed, by setting $g_0=0$ in Eqs.~(\ref{ltl0}), (\ref{gtg0}), (\ref{Zi}), and (\ref{mtm+}) 
[the latter being needed only to $O\left(h^2\right)$], we can obtain power series for 
$\l(\l_0,m)$ and $g(\l_0,m)$. Inverting the former to have $\l_0(\l,m)$
and substituting in the latter gives:
\begin{equation}
g = \frac{9 h \lambda ^3}{\pi  m^3}
\left(1-\frac{3 h \lambda }{\pi  m}+
\frac{h^2 \lambda ^2 \left(18 {k_3}-24 {k_4}-274-27 \log
   \left(\frac{4}{3}\right)\right)}{3 \pi ^2 m^2}\right)
   +O\left(h^4\right).
\label{glm}
\end{equation}
The numerical value of the last coefficient, namely, 
$$
\frac{18 {k_3}-24 {k_4}-274-27 \log \left(\frac{4}{3}\right)}{3 \pi ^2 } = 1.38996295137,
$$
perfectly matches the one obtained by Sokolov et al with other methods 
\cite[eq. 2.5]{Soko-UO} or \cite[eq. 8]{Soko-Kud}.

In an analogous way, we can make $\l_0=0$ and obtain power series for 
$\l(g_0,m)$ and $g(g_0,m)$, to derive from them the function $\l(g,m)$:
\begin{equation}
\l = -\frac{15 \,h \,g\, m}{4 \pi }
\left(
1-\frac{45 h^2 g  \left(5- 2\log \left(\frac{4}{3}\right)\right)}{2 \pi^2}
\right)
   +O\left(h^4\right)
\label{lgm}
\end{equation}
[without $O\left(h^2\right)$ term]. This result, corresponding to the 
pure sextic theory $(g\phi^6)_3$, can be related to previous results 
for this theory \cite{Hager,Zin-Cod,AKT,McK,HT,Huish,Shrock,Kharuk}. 
We shall reveal the relationship in Sect.~\ref{separatrices}. 

Finally, it is worth noting that our regularization scheme reduces the non-universality to
logarithmic terms, specifically, 
a $(-\log\L)$ term (within $A$) and a $\log m$ term which, together, form 
a $\log (m/\L)$ term.
It is important to point out that the appearance of logarithmic terms, due to the marginal variable, 
with the consequent violation of universality,  
was noted long ago \cite{Wegner-R,Stephen,Law-Sar}. 
However, we will see that the beta functions are universal and that {\em all} the 
non-universal terms can be recovered by integrating these functions.



\section{Renormalization group}
\label{RG}

The renormalized effective field theory is characterized by three parameters, namely, 
the mass and the two coupling constants. Assuming that $m \ll \L$ and the scale $\L$ 
has practically disappeared from the theory, we have to refer the theory to the remaining scale 
$m$ (the correlation length scale). One pair of coupling constants can be related to another 
by a change of mass scale; that is to say, we can understand two instances of the 
field theory as essentially describing the same phenomena observed at different scales. 
This is what the renormalization group achieves. Thus, the RG action is specified 
by how the pair of coupling constants change with the scale $m$ for 
given bare coupling constants. 

We have already focused on a specific bare theory, the simple quartic theory. This 
theory is given by the functions $\l(\l_0,m)$ and $g(\l_0,m)$, where we assume that 
$\l=\l_0$ for an ``initial'' value of $m=m_0$ (to be identified with the bare mass in 
Sect.~\ref{genRG}). 
Thus, we understand the couple of functions 
in the space $(\l,g)$ as describing a curve parametrized by $m$ and passing by a point 
$(\l_0,g_0)$. This is a particular RG trajectory, which is especially important. 
Actually, to achieve an adequate formulation of the RG, we need to use 
dimensionless coupling constants, namely, $u=\l/m$ and $g$. Thus, the particular RG trajectory 
in the space $(u,g)$ that corresponds to the quartic theory is directly read 
from Eq.~(\ref{glm})  (to 3-loop order, of course). The special role of 
this RG trajectory will be discussed below. 

The perturbative renormalization group not only relates the physics at different scales but also 
provides extra information, as we shall see.
First of all, we shall obtain the RG beta functions.

\subsection{Perturbative RG for $(\lambda\phi^4+g\phi^6)_3$ theory}
\label{RGbetas}

The RG equations are differential equations that give the change of the pair 
$(u,g)$ with a change of $m$ for fixed $(\l_0,g_0)$ (fixed bare potential). 
They constitute an autonomous system of differential equations when 
we take as independent variable the ``RG time'' $\tau = \log(m_0/m)$ 
($m_0$ being now an arbitrary ``initial'' value) \cite{I}.
Thus, they define a vector field in the space $(u,g)$, whose components are the 
RG beta functions.

Our beta functions, which assume the physical mass $m$ as the relevant scale, 
are similar to Sokolov's \cite{Soko} (see also \cite{Goro}). In contrast, 
there have been later calculations of beta functions for the sextic theory 
that employ a different scale, mostly, 
the mass scale $\mu$ introduced in dimensional regularization 
\cite{McK,HT,Huish,Hager,Zin-Cod,Shrock,AKT}. 
As this procedure dismisses finite but mass-dependent terms in the renormalization process, 
the corresponding beta function are only valid in a restricted region of the space $(u,g)$ 
(named ``tricritical region'' by Sokolov \cite{Soko}). We will discuss this question at length
below. 

We can calculate the three-loop perturbative beta functions for dimensionless coupling constants 
$u=\l/m$ and $g$ from Eqs.~(\ref{ltl0}), (\ref{gtg0}), and (\ref{Zi}). They are obtained 
in appendix \ref{betas} and read: 
\begin{align}
\b_1 &= m\left(\frac{\p u}{\p m}\right)_{\!\!\l_0,\,g_0} =
-u-\frac{3 h \left(5 g-6 u^2\right)}{4 \pi }
+\frac{h^2}{4\pi^2}\left(\frac{907 u^3}{9}-\frac{165 g u}{2} \right)
   \nonumber\\
   &+ \frac{h^3}{48\pi^3} \left\{-8100 \,g^2 \log \left(\frac{4}{3}\right)
   -5 g u^2 \left(240 {k_3}-\Red{1283}-1404 \log
   \left(\frac{4}{3}\right)\right)
   \right.\nonumber\\
   & \left.\phantom{==}
   {}+ 6 u^4 \left(144 {k_3}-96 {k_4}-24{k_5}-911-324 \log
   \left(\frac{4}{3}\right)\right)
   \right\},
\label{b1}
\\
\b_2 &= m\left(\frac{\p g}{\p m}\right)_{\!\!\l_0,\,g_0} =
\frac{9 h \left(5 g u-6 u^3\right)}{2 \pi}
+\frac{h^2}{\pi^2}\left(\frac{1275 g^2}{8} -\frac{2557 g u^2}{12}+ 27 u^4\right) 
   \nonumber\\
   &+ \frac{h^3}{8\pi^3} \left\{-5 g^2 u \left(280 {k_3}-\Red{2119}-2160 \log
   \left(\frac{4}{3}\right)\right)
   \right.\nonumber\\&\phantom{==} \left.
   {}- g u^3 \left(-3960 {k_3}+1920 {k_4}+36{k_5}+63797+11340 \log \left(\frac{4}{3}\right)\right)
   \right.\nonumber\\
   &\phantom{==} \left.
   {}- {6 u^5 \left(360 {k_3}-480 {k_4}-7891-540 \log
   \left(\frac{4}{3}\right)\right)}
   \right\}.
\label{b2}
\end{align}
These differential 
equations are scheme-independent, since they have no trace of scheme-dependent constants. 
We have kept $h$ to distinguish the contributions of several orders but 
we shall set $h=1$ to plot the RG flow. 

Naturally, the beta functions~(\ref{b1}) and (\ref{b2}) can be compared with previous calculations, 
especially those in Refs.~\onlinecite{Goro,Soko}. 
Unfortunately, these calculations are not consistent with respect 
to perturbative order (powers of $h$). Therefore, some terms of Eqs.~(\ref{b1}) and (\ref{b2}) appear 
in them, but others do not. Furthermore, those two calculations are not consistent with each other 
and are therefore hardly useful.
 
On the other hand, we can compare the beta functions~(\ref{b1}) and (\ref{b2}) with our preceding 
two-loop calculation \cite{I}. 
Since the beta functions are scheme-independent, one might have expected that 
the expressions in Ref.~\onlinecite{I} should have been {\em exactly} recovered. 
However, the comparison reveals small differences in some two-loop coefficients. 
Of course, these differences are due to having considered here the field renormalization, 
unlike in Ref.~\onlinecite{I}. There, $Z=1$ was assumed, for simplicity, arguing that it was 
so close to one that it would not make a sizable change. Indeed, the coefficients hardly change.

\begin{figure}
\includegraphics[width=10cm]{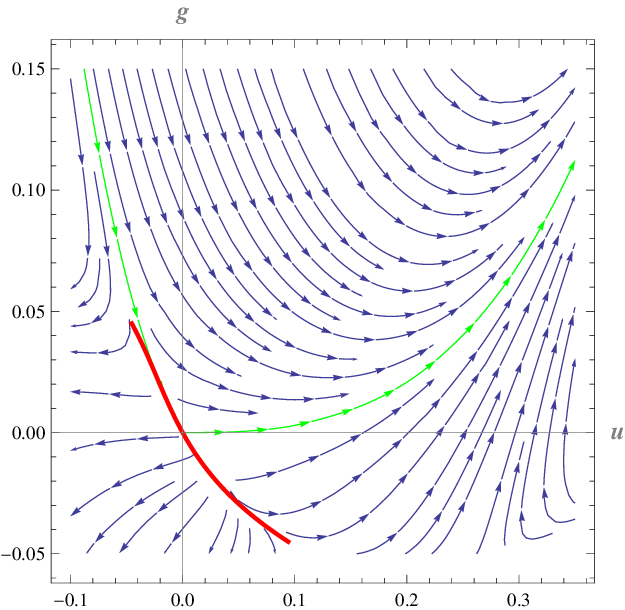}
\caption{RG flow for decreasing mass, with its separatrices (in green). The three-loop 
polynomial approximation to the second separatrix is highlighted (in red).}	
\label{fig1}
\end{figure}

The RG flow is plotted in Fig.~\ref{fig1} (with other features to be explained below). 
Of course, this plot is similar to the one in Ref.~\onlinecite{I}, because the magnitude of 
the 3-loop contribution to the beta functions is small near the origin 
(the tricritical fixed point). The 3-loop contribution grows as we consider the flow farther 
from the origin, but perturbation theory also becomes less accurate. In fact, 
the loop expansion is asymptotic and we should consider where the expansion becomes 
unreliable. As in any asymptotic expansion, the addition of new terms improves the accuracy 
of the expansion near the expansion point but restrict its domain of validity. 
Since we are chiefly concerned with the tricritical to critical crossover, 
which can be studied in the vicinity of the tricritical fixed point, 
the 3-loop contribution is useful.

\subsubsection{Note on previous beta functions calculated using dimensional regularization}

Let us quote the beta functions obtained with dimensional regularization ($\ve=3-d$)
\cite{McK,HT,Huish,Zin-Cod}, up to the two-loop order:
\begin{align}
\mu\frac{d m^2}{d \mu} &= -2 m^2 + h^2\frac{6\l^2}{\pi^2}, \label{m2mu}\\
\mu\frac{d \l}{d \mu} &= -\left(1+\ve - h^2\frac{30 g}{\pi^2}\right)\l, \label{lmu}\\
\mu\frac{d g}{d \mu} &= -2\ve g + h^2\frac{75\,g^2}{\pi^2}.
\label{gmu}
\end{align}
(the linear terms are missing in Refs.~\onlinecite{McK,HT,Huish}). 

Essentially equivalent equations were already found by Stephen et al \cite{Stephen}, in their 
analysis of logarithmic corrections to the mean-field theory of symmetrical tricritical points, 
using what they called a ``graphical'' method. 
Beta functiona for the two coupling constants, 
equivalent to Eqs~(\ref{lmu},\ref{gmu}), were also obtained 
in Ref.~\onlinecite{Carvalho}, in the {\em deep Euclidean region}.
Lawrie and Sarbach's beta functions \cite{Law-Sar}
also contain the terms present in Eqs.~(\ref{m2mu},\ref{lmu},\ref{gmu}).   
We are cautioned that those beta functions
are ``restricted to studying tricritical singularities.'' 

In $d=3$, the beta functions (\ref{m2mu},\ref{lmu},\ref{gmu}) are certainly different from our 
beta functions~(\ref{b1}) and (\ref{b2}). In fact, equations (\ref{m2mu},\ref{lmu},\ref{gmu}) can be 
successively integrated, in reverse order \cite{Stephen,Law-Sar}  
(the last equation gives $g(\mu)$, and  
expressions in terms of mass result from the first equation).
Equations~(\ref{m2mu},\ref{lmu},\ref{gmu}), 
as well as their higher-order versions, can be applied to the tricritical 
fixed point, which becomes non trivial in $d<3$. 
However, they are incapable of describing the crossover in $d=3$: It is apparent that they do not  
have a non-trivial fixed point, unlike Eqs.~(\ref{b1}) and (\ref{b2}). Indeed, 
Lawrie and Sarbach write: ``To describe this crossover, a better approximation to
the scaling function in [\ldots] is required. At the time of writing,
such a function has not been obtained'' \cite{Law-Sar}.

\section{The renormalization group as a dynamical system}
\label{dynamical}

The general study of autonomous systems of differential equations belongs to 
the well developed theory of dynamical systems \cite{Gu-Ho,Perko}. 
Methods of this theory have been often employed to treat 
some aspects of the renormalization group. It was proposed very early that 
the RG could be a {\em gradient dynamical system}, which would imply simple dynamics 
and simple asymptotic behavior \cite{WZ}. RG monotonicity properties, related to the existence 
of a {\em Liapunov function} \cite{Gu-Ho,Perko}, were first proved in two-dimensional 
field theory \cite{Zamo} and subsequently generalized 
\cite{Jack-O,J-OC,JG,Apenko,Shore}.
Morozov and Niemi summarize several efforts in this regard and discuss further 
aspects of the question (the possibility of chaotic RG flows, etc) \cite{Morozov}.
The connection with bifurcation theory and topological methods is studied by Gukov 
\cite{Gukov}.

Regarding our RG equations, we only need 
the theory of planar dynamical systems, which is simpler and is extensively studied 
\cite{Andronov,Llibre}. The most important feature of our beta functions is their singular point 
at the origin, which is the tricritical fixed point. 
As we shall see, it is of {\em semi-hyperbolic} type and, in fact, it is a {\em saddle-node} 
singular point \cite{Andronov,Gu-Ho,Perko,Llibre}. 

\subsection{Nature of the tricritical fixed point}

The study of a singular point of an autonomous differential system (or of its defining vector field) 
begins with characterizing its linear part at that point. Some comments on the 
linear beta functions were made in Ref.~\onlinecite{I}. The standard procedure is 
to analyze the Jacobian matrix (the matrix of partial derivatives) at the singular point, 
which is a $2 \times 2$ matrix in the planar case \cite{Andronov,Gu-Ho,Perko,Llibre}. 
The simplest possibility is that the matrix is regular, but the case of null eigenvalues 
is also thoroughly studied. 
The regular case is called {\em hyperbolic}. 
When just one eigenvalue is null, the name is {\em semi-hyperbolic}. A 
singular point of this type can be the origin of multiple {\em separatrices}, which are
flow lines that originate from the point and divide its neighborhood into a finite
number of open regions called sectors. In each sector, the flow is qualitatively different. 
The separatrices of the flow in a neighborhood of the tricritical point are easily identified 
and have been briefly analyzed in Ref.~\onlinecite{I}.

In fact, it is easy to apply the standard procedure to our beta functions, 
Eqs.~(\ref{b1}) and (\ref{b2}) (with $h=1$). At the origin, 
the Jacobian matrix eigenvalues are $(-1,0)$, so it is a semi-hyperbolic singular point. 
Therefore, the analysis of the linear part is not sufficient. 
For a semi-hyperbolic singular point, we can apply to the nonlinear part 
``Theorem 65'', proved by Andronov et al \cite[p.\ 340]{Andronov} 
(also \cite[sec.\ 2.6]{Llibre}). Thus we find that 
the tricritical point is a {\em saddle-node}, namely, a semi-hyperbolic singular point 
whose neighborhood consists of two hyperbolic sectors and one
parabolic sector, with three separatrices. The two hyperbolic sectors are placed 
in the half-plane $g>0$, whereas the parabolic sector is mostly placed 
in the half-plane $g<0$ (Fig.~\ref{fig1}). Naturally, this occurs because 
two separatrices are tangent to the $u$-axis. 

Let us comment on the name {\em saddle-node}. This name 
is used in bifurcation theory, referring to the simplest {\em normal form} 
\cite[sec.\ 3.4]{Gu-Ho}. In a planar system that depends on a parameter and has one 
regular saddle point and one regular node, a bifurcation takes place when they merge 
for some value of the parameter. Then, the {\em phase portrait} (the geometry of the flow)
is precisely of the saddle-node type. Nonetheless, this same phase portrait can arise in other 
types of bifurcation, for example, in the {\em transcritical bifurcation} \cite[sec.\ 3.4]{Gu-Ho}.  
This type is indeed the one corresponding to the tricritical fixed point when we consider the space 
dimension as a bifurcation parameter (as in the $\ve$ expansion). However, we are not concerned here 
with space dimension variations but we are considering fixed $d=3$.

Before geting into any calculation of the features of the phase portrait about the tricritical point, 
let us introduce some more concepts of dynamical system theory. Semi-hyperbolic singular points 
are more complex than their regular counterparts of the hyperbolic type. In general systems, 
the presence of null eigenvalues thoroughly alters the exponentially 
contracting-expanding dynamics of hyperbolic flows. 
In a planar system, along the separatrix corresponding to the null eigenvalue, 
the dynamics is not exponential but slower. 
If the other eigenvalue is negative, the dynamics is exponentially attracted to the 
slower flow on the separatrix. However, it should be noted that the sign of that eigenvalue 
obviously depends on the sign of the time variable, and we actually prefer the opposite flow, as 
in Fig.~\ref{fig1}.

In general, with the choice of negative signs for non-zero eigenvalues, since the dynamics are 
exponentially attracted to a lower-dimensional manifold, the qualitative behavior in a neighborhood 
of the singular point is determined by its behavior on that manifold, which is called the {\em center 
manifold} \cite{Gu-Ho,Perko}. At a saddle-node point, we have a center line that partially coincides 
with the separatrix corresponding to the zero eigenvalue. Only partially, because the center line can 
be prolonged through the singular point, along the direction determined by the given eigenvector. 
However, this prolongation is not unique, although there can be only one analytic center manifold 
\cite{Gu-Ho,Perko}. Now, let us calculate the separatrices and the center line in our case, under the 
assumption of analyticity or, at least, smoothness.

\subsection{RG flow and separatrices}
\label{separatrices}

The separatrix tangent to the $u$-axis, for $u>0$, has been studied in Ref.~\onlinecite{I}, 
under the name of ``connecting line''. This name is due to the natural assumption that it leads 
to the non-trivial RG fixed point (the Wilson-Fisher fixed point in $d=3$). However, 
let us remark that a proof of it  
falls outside the local analysis we are now considering. 
Whether it is a connecting line or just a separatrix, 
the local analysis is the same and is based on 
assuming a smooth solution to the single differential equation associated to the autonomoous system:
$$\frac{dg}{du} = \frac{\b_2(u,g)}{\b_1(u,g)}\,.$$
To wit, the solution $g(u)$ must be such that $g(0)=0$ and have an expansion in powers of $u$  
(possibly, an asymptotic expansion, if $g(u)$ is not analytic at $u=0$). 
Naturally, the two-loop expansion in Ref.~\onlinecite{I} is still valid, and we further find:
\begin{equation}
g(u) = \frac{9 u^3}{\pi }  \left(1-\frac{3 u}{\pi } +1.38996 u^2 + 1.79909 u^3 + O\left(u^4\right)\right).
\label{gu3loop}
\end{equation}
(The numerical values of $k_3$, $k_4$, and $k_5$ have been substituted for.) 
This expression agrees with Eq.~(\ref{glm}) and adds one order. Unfortunately, 
this RG ``ìmprovement'' is of little use, since even its \Red{sign} is wrong, 
when compared to the known five-loop expansion of $g(u)$ \cite[Eq 8]{Soko-Kud}.

As regards the tricritical-critical crossover, the second separatrix is even more interesting, 
because it separates the (hyperbolic) sector that flows towards positive $u$, and presumably towards 
the non-trivial fixed point, from the sector that eventually flows towards negative $u$. 
Moreover, this separatrix coincides with the center line, where the flow has a special character. 
Naturally, the separatrix is not tangent to a coordinate axis but to the direction of 
the corresponding eigenvector, given by the equation
$$
v=u+\frac{15 g h}{4 \pi }=0.
$$
Therefore, we now seek a smooth solution to another differential equation:
$$\frac{dv}{dg} = \frac{\b_1(u,g)}{\b_2(u,g)}+\frac{15 h}{4 \pi }\,.$$
That is to say, we assume a solution $v(g)$ such that $v(0)=0$ and such that it has an expansion 
in powers of $g$. 
Matching coefficients, we find:
\begin{align}
u(g) = v(g) - \frac{15 g h}{4 \pi } &= -\frac{15 g h}{4 \pi } 
   + \frac{675 \,h^3 g^2 \left(5 - 2 \log\left(\frac{4}{3}\right)\right)}{8 \pi ^3}  \nonumber\\
   &+ \frac{375 \,g^3 h^5 \left(360 {k_3}-\Red{12731}+567 \log
   \left(\frac{4}{3}\right)\right)}{64 \pi ^5}+O\left(h^7\right).
\end{align}
Here, $O\left(h^7\right)$ can be replaced with the equivalent order $O\left(g^5\right)$. 
We explicitly show the powers of $h$ to better keep track of the perturbative order. 
The obtained expression agrees with Eq.~(\ref{lgm}) for the pure sextic
theory (and adds one order).
Of course, the 3-loop effective potential only guarantees exactness up to $O\left(h^3\right)$, 
while the $O\left(h^5\right)$ term is a guess (or an RG improvement).
This polynomial approximation is highlighted in Fig.~\ref{fig1}.

The next step is to find the respective one-dimensional RG flows on the separatrices. 
On the first one, we have:
\begin{align}
\b_1[u,g(u)] &= 
-u+\frac{9 u^2}{2 \pi }-\frac{77 u^3}{9 \pi ^2}+\frac{u^4
   \left(72 {k_3}-48 {k_4}-12 {k_5}-793-162 \log
   \left(\frac{4}{3}\right)\right)}{4 \pi ^3}+O\left(u^5\right)
   \nonumber\\
   &= -u+\frac{9 u^2}{2 \pi }-\frac{77 u^3}{9 \pi ^2}+ 1.03177\,u^4 +O\left(u^5\right).
\label{beta4}
\end{align}
It can be verified with the known $\b$-function of the quartic model, Eq.~(\ref{betaf4}).
We could also calculate 
$\b_1[u,g(u)+\d g]$ and $\b_2[u,g(u)+\d g]$, in the first order of $\d g$. 
This approximation would constitute an improvement on Sokolov's {\em critical region} approximation 
\cite{Soko}.

On the second separatrix, we have:
\begin{align}
\b_2[u(g),g] &= 
\frac{75 g^2 h^2}{\pi ^2}+\frac{75 g^3 h^4 \left(140 {k_3} + \Red{630} -1890 \log
   \left(\frac{4}{3}\right)\right)}{16 \pi ^4}+O\left(g^4\right)
   \nonumber\\
   &= \frac{75 g^2 h^2}{\pi ^2} + \Red{67.2291} g^3 h^4 +O\left(g^4\right).
\label{beta6}
\end{align}
Given that this separatrix is also a center line, we have achieved, 
in the terminology of dynamical systems, a {\em center manifold reduction}, 
based on the center manifold theorem \cite{Gu-Ho,Perko}. 
We could further obtain the RG flow in Sokolov's {\em tricritical region}, which is 
in fact the region 
where the sextic field theory is usually considered 
(namely, where $\l_0^2/m^2 \ll g_0$).
Eq.~(\ref{beta6}) indeed agrees with the usual beta function of this theory, 
Eq.~(\ref{gmu}), calculated by taking the derivative with respect to the mass 
scale $\mu$ in dimensional regularization 
\cite{McK,HT,Huish,Hager,Zin-Cod,Shrock,AKT} (or by other methods \cite{Stephen,Carvalho,Law-Sar}). 
Of course, Eq.~(\ref{beta6}) only reproduces exactly 
the usual beta function up to the two-loop order, while 
the 4-loop order in Eq.~(\ref{beta6}) is a guess. 

The beta function of the sextic field theory has also been calculated for the $N$-vector model 
(in dimensional regularization) 
\cite{Hager,Zin-Cod,Shrock}. This generalization is interesting, because the theory can be solved 
in the large-$N$ limit, where a non-trivial fixed point exists \cite{Shrock}. Presumably, this 
point could survive for lower $N$ and even for $N=1$. In the terminology of dynamical systems, 
the center-line separatrix would then be a connecting line, like the first separatrix. 
However, the six-loop beta function calculated by Shrock \cite{Shrock} suggests that 
the non-trivial fixed point already disappears for fairly large $N$. 

Eqs.~(\ref{beta4}) and (\ref{beta6}) are easier to analyze, since they define one-dimensional flows. 
Of course, the flow of $u$ is exponential, while the flow of $g$ is not. If we reverse the flow, 
that is, we consider the flow towards the UV scales, the $u$-flow at the tricritical point 
is stable while the $g$-flow is not. 
In Fig.~\ref{fig1}, 
we observe that the IR $g$-flow leads to the tricritical point, while in the second separatrix, 
that is to say, while $g>0$. It does not seem that $g<0$ makes any physical sense, 
as long as it corrresponds to $g_0<0$ in the bare sextic potential, 
which is then unbounded below. At any rate, to study 
the physics of the crossover, we first need more information on the full two-dimensional RG flow.

\subsubsection{Note on the beta function of the sextic field theory}

Redefinitions of coupling constants are normaly considered as scheme transformations, 
usually in regard to four-dimensional scalar field theory \cite[\S 10.11]{ZJ}. The same 
argument is valid in three-dimensional scalar field theory and shows that 
the first two terms of Eq.~(\ref{beta6}) are universal in this sense, while the others 
are arbitrary and therefore can be made to vanish (this argument is 
used, for example, by Shrock \cite{Shrock}). In the terminology of dynamical systems,
the first two terms constitute the {\em normal form} of this saddle-node singular point 
\cite[sec.\ 2.6]{Llibre}. 

The differential equation (\ref{beta6}) restricted to the first term (that is, to the two-loop 
order) is easily integrated and produces terms with $\log (m/\L)$. The normal form of the equation, 
with the first two terms, can also be integrated, in terms of the {\em Lambert W function}, 
as was found in the study of the QCD beta function, which adopts a similar form 
\cite{GGK}. The asymptotic expansion of the Lambert function is well known 
and contains, in addition to powers of $\log (m/\L)$, powers of $\log|\log (m/\L)|$. 
Naturally, the latter have smaller magnitude. 

It is also remarkable the considerable effect that the first term of the beta function has by itself, 
when $m\ll \L$ \cite{Stephen,Carvalho}. Integrating it, one obtains:
\begin{equation}
g(m) = \frac{g_0}{1 - 75 h^2g_0\log (m/\L)/\pi^2}\,.
\label{glog}
\end{equation}
If $m\lesssim \L$, then we can expand it in powers of $h^2$, matching  
Eq.~(\ref{gtg0}) [or Eq.~(\ref{grg0})] up to $O(h^2)$, provided that $\l_0=0$. At the same time,  
higher-order terms appear (as a RG improvement).  
In contrast, when $m\ll \L$, $g$ becomes independent of $g_0$ and vanishes as 
$\left|\log (m/\L)\right|^{-1}$. Of course, any $\l_0\neq 0$ prevents this vanishing and 
induces a crossover towards the non-trivial fixed point.

\section{General features of the RG flow}
\label{genRG}

In  terms of the ``RG time'' $\tau = \log(m_0/m)$, the beta functions~(\ref{b1}) and (\ref{b2}) 
constitute an autonomous system, but a highly nonlinear one. Thus, we cannot expect to find 
analytical solutions; except in perturbation theory, of course. Fortunately, we do not have to 
worry about how to find perturbative solutions, because the general solution of the system, 
up to the present order, is given by 
Eqs.~(\ref{ltl0}), (\ref{gtg0}), (\ref{Zi}), and (\ref{mtm+})
[that is, by Eqs.~(\ref{lrl0}) and (\ref{grg0})].
The two arbitrary constants in those equations, namely, $\l_0$ and $g_0$, 
which had the meaning of bare coupling constants, need to be understood 
as integration constants, since they are kept constant in the derivation of beta functions 
(these are functions of renormalized coupling constants only). 
Along the way, we can give a new meaning to the scheme dependence. The rationale has been introduced 
in Ref.~\onlinecite{I}.

The case of the super-renormalizable quartic theory ($g_0=0$) is simpler. 
With $g_0=0$, the renormalization equation~(\ref{ltl0}) is scheme independent. 
Furthermore, the limit $m\ra\infty$ is feasible and makes $\l \ra \l_0$. 
On the other hand, in the effective theory, one must assume $m \ll \L$, so we can interpret 
the limit $m\ra\infty$ 
as actually meaning $m \ra \L$. 
Therefore, this allows us to interpret the bare coupling as $\l_0=\l(m \simeq \L)$. 
Of course, this interpretation is supported by 
calculations with the exact RG, where one starts with bare coupling constants 
at finite cutoff $\L$ 
and integrates it down to $\L=0$ 
\cite{I}.

The renormalizable sextic theory is subtler, because the limit $m\ra\infty$ is not 
feasible in Eqs.~(\ref{ltl0}) and (\ref{gtg0}). Nevertheless, we can again interpret this limit as 
the limit $m \ra \L$, assuming that $\l_0=\l(m \simeq \L)$ and $g_0=g(m \simeq \L)$. 
However, these ``bare'' constants cannot now be identified with the two arbitrary 
constants $\l_0$ and $g_0$,
which are just two arbitrary
integration constants of the RG flow, and which hence we rename as $\widetilde{\l}_0$ and 
$\widetilde{g}_0$ to distinguish them from the newly defined $\l_0$ and $g_0$. 
To relate the pair $(\l,g)$ at an arbitrary $m \ll \L$ to 
the pair $(\l_0,g_0)$ at $m \simeq \L$, we need to get rid of the integration constants 
$\widetilde{\l}_0$ and $\widetilde{g}_0$. That is to say, we have to trade 
integration constants for initial conditions, as in any dynamical problem. Let us see how.

Naturally, the constants $\widetilde{\l}_0$ and $\widetilde{g}_0$ can be obtained from 
Eqs.~(\ref{ltl0}) and (\ref{gtg0}) by inverting the power series. 
Once the inversion is made, it is easy to get rid of $\widetilde{\l}_0$ and $\widetilde{g}_0$. 
Indeed, we only have to make 
$$\widetilde{\l}_0(\l_1,g_1,m_1)=\widetilde{\l}_0(\l_2,g_2,m_2), \quad \widetilde{g}_0(\l_1,g_1,m_1)=\widetilde{g}_0(\l_2,g_2,m_2)$$ 
to obtain the relation between two arbitrary sets of renormalized parameters. Then, we can keep 
one set arbitrary, as $(\l,g,m)$, and assign the other to $(\l_0,g_0,\L)$. Of course, this 
assignation does not break the symmetry between the two sets, which implies that we obtain 
inverse powers of $\L$ as well as inverse powers of $m$ [the latter being expected, since 
they are present in Eqs.~(\ref{ltl0}) and (\ref{gtg0})]. 
Nevertheless, we can suppress the inverse powers of $\L$, assuming small values of 
the dimensional parameters, despite 
losing a strict interpretation of $(\l_0,g_0)$ as the initial condition.

The complete calculation to $O\left(h^3\right)$ is cumbersome. Therefore, we just display 
the results up to $O\left(h^2\right)$:
\begin{align}
\l &= \l_0
+ h \left(
\frac{15 g_0 }{4 \pi }\, \Lambda - \frac{15 g_0 m}{4 \pi }-\frac{9 \lambda _0^2}{2 \pi  m}
\right)
\nonumber\\
&- h^2 \left(
\frac{135 g_0 \lambda _0 }{4 \pi ^2 m}\,\Lambda -\frac{30 g_0
   \lambda _0 }{\pi ^2}\,\log \frac{m}{\L} -\frac{945 g_0
   \lambda _0}{8 \pi ^2}-\frac{883 \lambda _0^3}{36 \pi ^2 m^2}
\right)
   +O\left(\frac{1}{\Lambda}\right) +O\left(h^3\right).
\label{ll0L}
\end{align}
\begin{align}
g &= g_0
+ h \left(
\frac{9 \lambda _0^3}{\pi  m^3}-\frac{45 g_0 \lambda _0}{2 \pi  m}    \right) 
+ h^2 \left\{
135 \left(
\frac{3 g_0 \lambda _0^2}{4 \pi ^2 m^3}-\frac{5 g_0^2}{8 \pi ^2 m}\right)\Lambda \right.
\nonumber\\
&\left.
 {}+ \frac{75 g_0^2}{\pi ^2}\,\log \frac{m}{\L}
 +\frac{675 g_0^2}{8 \pi ^2}
 +\frac{3101 g_0 \lambda _0^2}{12 \pi ^2 m^2}
 -\frac{297\lambda _0^4}{2 \pi ^2 m^4}
   \right\} +O\left(\frac{1}{\Lambda}\right) + O\left(h^3\right).
\label{gg0L}
\end{align}
These two expressions reproduce, up to $O\left(h^2\right)$, 
the expressions of $\l$ and $g$ obtained from Eqs.~(\ref{ltl0}) and (\ref{gtg0}), namely, 
Eqs.~(\ref{lrl0}) and (\ref{grg0}),
but they have extra terms, proportional to $\L$. Of course, the scheme dependent constant
$A$ in Eqs.~(\ref{lrl0}) and (\ref{grg0}) cannot be reproduced exactly. 
We must consider that we have made a sort of scheme choice with the assignation of initial 
conditions at $(\l_0,g_0,\L)$. This choice can be somewhat different,  
because the initial condition has been loosely defined,  
at $m \simeq \L$, which allows for a number of the order of unity. 



In the result of the full calculation up to $O\left(h^3\right)$, more positive powers of $\L$ 
appear, and still more in higher orders 
(some of them are shown in Ref.~\onlinecite{Kharuk}). 
We can appreciate the advantage of schemes that avoid them, either through 
dimensional regularization or as we have done. We can always 
recover them {\em a posteriori}.

We must also remark the presence of inverse powers of $\L$ in Eqs.~(\ref{ll0L}) and (\ref{gg0L}), 
despite the fact that they can be suppressed, 
and we may wonder whether they could have some role. As noticed above, these terms realize 
the symmetry between $(\l_0,g_0,\L)$ and $(\l,g,m)$ in the RG. As much as we can 
obtain the beta functions by taking derivatives with respect to $m$ for fixed 
$(\l_0,g_0,\L)$, we can as well obtain them by taking derivatives with respect to $\L$ 
for fixed $(\l,g,m)$. This can be useful as an alternative method, related to 
the method of ``bare RG equations'' \cite[\S 25.4]{ZJ}.

\subsubsection{Note on the comparison with the exact RG}
\label{ERG}

The comparison between two-loop perturbative results and exact RG integrations, carried out 
in Ref.~\onlinecite{I}, is successful, that is to say, good agreement is found where expected. 
Of course, better agreement is found for small $g$ and $u$ (in fact, a quite small value of $g_0$ 
is actually required for perturbation theory to work). Relatively good agreement is found up to, 
say, $u=0.2$ (corresponding to $m/\L=0.04$). For this value, the difference in $u$ 
between perturbative and exact RG results is $<0.3\%$, 
although the difference in $g$ is about $18\%$ \cite{I}. 

Let us try to assess how good the agreement is, using a simplified method. 
The accuracy of perturbation theory for small $g_0$, that is, near the connecting line 
$g(u)$ of Eq.~(\ref{gu3loop}), 
can be estimated as follows. Let us assume that most of the error is perturbative 
(even though the formulation of the exact RG employed in Ref.~\onlinecite{I} is 
not entirely ``exact''). Therefore, we estimate the error in $u$ from Eq.~(\ref{l0lf4}).
For $u=0.2$, we can estimate the error in the two-loop calculation by the three-loop term, obtaining 
$1.8\cdot 0.2^3 \simeq 0.01$, which is larger than the actual error. 
On the other hand, we estimate the error in $g$ from 
Eq.~(\ref{gu3loop}), obtaining $1.4\cdot 0.2^2 \simeq 0.06$. In this case, it is considerably 
smaller than the actual error. 

Since the three-loop order cannot account for the discrepancy in the values of $g$, we could 
infer that most of the error is not perturbative. That is, 
the problem should lie in the formulation of the exact RG employed in Ref.~\onlinecite{I}, 
specifically, in the truncation made. 
The number of coupling constants used in the effective potential approximation is sufficient, 
because increasing this number produces totally negligible changes. In conclusion, the discrepancy 
probably lies in the effective potential approximation, which discards derivative couplings. 

\subsection{Crossover behaviour}

In crossover critical behaviour, there is a gradual change of universality class. 
In a RG description, the change is due to the competition between fixed points 
\cite{Pfeu-T}. 
Indeed, a RG trajectory can pass close by one fixed point before
finally converging on another. This change is reflected in the successive appearance of two
different types of critical behaviour. 

The tricritical to critical crossover is a particular but common type of crossover behaviour, 
observed in systems with two intensive variables, at least, and a phase diagram 
where a line of first-order phase transitions changes into a line of second-order (continuous) critical transitions at a tricritical point. This name refers to the character of the 
point in a three-variable phase diagram, the additional variable being a  
symmetry-breaking field (for example, a sort of magnetic field conjugate to $\f$). In this 
phase diagram, there are two further critical lines that meet with the first one at the 
tricritical point, forming a typical wing structure together with the corresponding surfaces 
of first-order transitions \cite[Ch.\ 12]{Pfeu-T}, \cite{Law-Sar}.
Naturally, the crossover occurs near the tricritical point and involves the 
symmetric critical line.

The phenomenology of the tricritical to critical crossover is easily realized by the 
classical sextic potential (\ref{pot}) (the Landau theory) \cite[Sec.\ III]{Law-Sar}.
The symmetric variables are $m_0$ and $\l_0$ (one can set $g_0$ equal to a non-vaishing constant, 
for instance, $g_0=1$, by rescaling $\f$).  
We could consider the symmetry-breaking variable $H$, adding the 
term $H\phi$ to $U(\f)$, in order to have the complete phase diagram \cite[Sec.\ III]{Law-Sar}. 
In the symmetry plane of the phase diagram ($H=0$), we have a triple line of first-order phase 
transitions, where $m_0>0$ and $\l_0<0$, 
so that $U(\f)$ has three minima with the same value of $U$. This line ends at 
a tricritical point, with $m_0=\l_0=0$, where the three minima merge. 
One can prolong further the triple line beyond the tricritical point, 
where it changes into a line such that $m_0=0$ but $\l_0>0$, that is to say, 
an ordinary critical line. 

Of course, the effective potential is not so simple. However, we still have (for $m\neq 0$):
$$
U_\mathrm{eff}(\f) = \frac{m^2}{2}\,\f^2 + \l\,\f^4 + g\,\f^6 + O\left(\f^8\right).
$$
Furthermore, we can still identify the tricritical and critical points as points such that $m=0$. 
In the RG flow diagram (Fig.~\ref{fig1}), points such that $m=0$ are fixed points; namely, 
the origin and also the (presumably present) Wilson-Fisher fixed point. That is, 
fixed points necessarily correspond to $m=0$ or $m=\infty$ (RG time $\tau=\pm\infty$). 
Therefore, a fixed point with stable and unstable directions corresponds to both $m=0$ and $m=\infty$. 
Indeed, the origin in Fig.~\ref{fig1} is the end of the second separatrix and the beginning of 
the first one. 

The change of behaviour due to the competition between the two fixed points takes place near 
the origin, for RG trajectories in the first sector that begin close to the second separatrix, 
with small values of 
$\widetilde{\l}_0$ and $\widetilde{g}_0$ (bare couplings). 
An example has been analyzed in Ref.~\onlinecite{I}. 
It is not crucial that $\widetilde{g}_0$ be very small, but only that $\widetilde{\l}_0$ be, 
and thus the RG trajectory be close to the second separatrix and approach the origin 
(recall that $\widetilde{\l}_0=0$ at this separatrix). 
For a long interval of $\tau$, the tricritical point 
dominates, but eventually the trajectory turns towards the connecting line leading to 
the Wilson-Fisher fixed point. However, $\l$ hardly changes while $u$ grows \cite{I}, 
demonstrating the dominance of the tricritical point. In fact, $\l$ is stationary
even on the connecting line, provided that $m>\widetilde{\l}_0$, which we assume to be very small. 
This is a simple consequence of Eq.~(\ref{l0lf4}). In any case, 
$g$ becomes irrelevant on the connecting line, since it is a function of $\l/m$. 
Of course, many RG trajectories perform a crossover. 
One can be selected by setting 
$g_0$ (or $\widetilde{g}_0$) 
to a small value, for consistency with perturbation theory	 (not to $g_0=1$).

An important number associated to any crossover critical behaviour is the crossover exponent 
\cite{Pfeu-T}. 
For the tricritical to critical crossover, 
it was initially usually defined in terms of scaling fields 
\cite{R-Wegner,R-Wegner_1}, \cite[\S 12.2]{Pfeu-T}, 
\cite{Law-Sar} (scaling fields are coupling parameters that transform 
linearly under the renormalization group). In the end, what matters is the dimensions of 
the relevant couplings at the tricritical point. Since it is a trivial fixed point, 
the naive (classical) dimensions are not altered. 
We can take as 
coupling parameters the coefficients in the potential, namely, $m^2$ and $\l$. 
Therefore, the crossover exponent is predicted to be $\phi_\mathrm{cross}=2$, 
because the ratio $(\l)^2/(m^2)$ is scale invariant \cite{Law-Sar}
(that is, in exactly three dimensions, with corrections in $d<3$ \cite{Hager}).
Experimental values are indeed close to this prediction, as shown by experiments on
$^3$He-$^4$He mixtures and magnetic systems 
\cite{Wolf}.

In contrast with $m^2$ or $\l$, $g$ is dimensionless. Actually,
the dimension ratios imply that characteristic lines, such as the triple line, 
are given by equations of the form $g \propto \l^2/m^2$. Lines of this form can also be 
significant in the RG flow diagram (Fig.~\ref{fig1}). In fact, it was argued in Ref.~\onlinecite{I} 
that the line $g=6u^2/5$, where the RG flow was found to turn upward, can be used as a convenient 
definition of crossover from tricritical to critical behavior. This line, defined 
at the two-loop order, changes somewhat at the present three-loop order. Nevertheless, 
the line still defines the behavior of the RG flow close to the origin. 
In absence of a more precise definition of crossover, 
this intuitive definition is surely sensible.

On the other hand, 
it could also be argued that the crossover is realized when the ordered states
with $\langle\f\rangle\neq 0$ disappear (become unstable). This conditions demands that $\l <0$ 
and gives a different line, 
namely, $g=2u^2/3$ \cite{Law-Sar}. 
The crossover exponent does not change, of course. 
However, let us note that the part $u<0$ of any line $g \propto u^2$ is in 
the sector that eventually flows towards more negative values of $u$, at least, for points 
in the perturbative domain, close to the origin in Fig.~\ref{fig1}.

\subsection{Exact RG and crossover}

Since the exact RG results basically agree with the perturbative results,  
as indicated above (note~\ref{ERG}), we can try to gather more information about the crossover itself
from exact RG integrations. 
The non-perturbative RG 
scale transformations are actually linked to a progressive removal of short-distance degrees of freedom 
and provide a different perspective (although this RG is often not entirely ``exact''). 

The first use of a non-perturbative RG for tricritical behavior was actually made 
by Riedel and Wegner at the beginning of the development of RG methods \cite{R-Wegner}. 
The type of RG procedure that they employed was an iterative procedure known as the approximate recursion relation (this procedure is explained by Wilson and Kogut \cite[\S 6]{Wil-Kog}). 
Riedel and Wegner found, starting from certain initial constants, that 
the recursion relation stabilizes on a three-well shape of the potential 
that asymptotically becomes flat \cite{R-Wegner}. 
They called this limit the ``Gaussian tricritical fixed point'' \cite{R-Wegner}. Naturally, 
this limit corresponds to a trivial potential. 
The coefficient that gives the 
approach to the limit vanishes according to a law that is analogous to Eq.~(\ref{glog}).
Unfortunately, Riedel and Wegner did not specify their initial constants. 

Instead of the approximate recursion relation, we employ, like in Ref.~\onlinecite{I}, 
the Wegner-Houghton differential equation for the effective potential, which is obtained using 
a simple sharp wave-number cutoff \cite{Wegner-H}. (The results of other 
cutoff schemes are not expected to differ significantly \cite{I}.) 
Although this non-perturbative formulation of the RG is stronger than the approximate recursion 
relation, it was developed a little later (the role of Wegner in all these 
developments is remarkable). 
We work with a truncation of the Wegner-Houghton equation that includes in the effective 
potential terms up to $\phi^{16}$, and we put a sextic initial potential, as in Eq.~(\ref{pot}). 

The eight beta functions of the truncated Wegner-Houghton RG constitute a dynamical system, 
albeit a considerably more complex one than the planar system studied in 
Sec.~\ref{dynamical}, despite being related to it (note that the 
first ``coupling constant'' is now $m^2$). In any case, the standard procedure 
also requires the analysis of the Jacobian matrix at the singular point (the origin). 
This matrix is upper triangular, and therefore its eigenvalues are the diagonal terms of the matrix. 
The first two eigenvalues are positive (corresponding to $m^2$ and $\l$) and the third one is zero 
(corresponding to $g$), while the others are negative (corresponding to irrelevant couplings). 
We have again a semi-hyperbolic singular point. However, the center line can be 
much more complex in higher dimensions than in planar systems. In any case, 
the eigenvalues and eigenvectors still provide useful information about the dynamics 
near the singular point. 

The first eigenvector, with eigenvalue 2 (dimension of $m^2$), goes along the corresponding 
coordinate axis. The second eigenvector, with eigenvalue 1 (dimension of $\l$), 
is in the plane spanned by the first two axes, 
while the third eigenvector has components 
along the first three axes, namely,
$$\left(1,  -\frac{\pi^2}{3},\, \frac{2\pi^4}{45},\, 0, \ldots, 0\right).
$$
These two eigenvectors define a plane in the three-dimensional space of relevant 
or marginal couplings. 
This plane coincides with the plane identified in Ref.~\onlinecite[Sec.~5.1]{I} as tangent 
at the origin to the critical surface. 
Naturally, the character of the third eigenvector is marginal, in linear order, while that 
of the second eigenvector is relevant. A more detailed (nonlinear) analysis shows that 
the RG flow along the center half-line with positive $g$ leads to the origin, like in  
Sec.~\ref{dynamical}. Therefore, the flow over the critical surface is analogous 
to the flow in Fig.~\ref{fig1}.

Given that the potentials with parameters corresponding to points 
on the center half-line with positive $g$ are three-well shaped, and 
the flow along the line leads to the origin, this flow somehow  
reproduces the approach to the ``Gaussian tricritical fixed point'' observed by Riedel and Wegner 
\cite{R-Wegner}. 
However, we need a good deal of fine tuning to approach the origin, 
since there are two RG relevant parameters (a problem that Riedel and Wegner presumably also had). 
The crossover to the Wilson-Fisher fixed point also requires fine-tuning, because 
the most relevant parameter, $m^2$, tends to drive the flow 
away from the critical surface. 

At this point, it is worth noting that Riedel and Wegner casually mention (in an endnote) 
that they had to include a $\phi^{8}$ term in their initial potential \cite{R-Wegner}. 
We have not done so, because such a term is non-renormalizable in perturbation theory and we 
do not consider it here. Nevertheless, it can be included, because 
non-renormalizable terms are allowed in an effective field theory \cite[\S 12.4]{Weinberg}.
Surely, there may be some mathematical procedure that avoids excessive fine-tuning,  
with initial conditions in the three-dimensional parameter space, but this task is beyond 
the scope of this work.

\section{Discussion and Conclusions}

The sextic field theory has long been used for studying general properties of 
statistical models in ordinary three-dimensional space and, in particular, 
for studying the tricritical-critical crossover. Perturbative field theory is a suitable method 
for this purpose, but its full potential has not been exploited, for various reasons. 
A natural way of proceeding was shown in our preceding article, but that work was limited to a 
two-loop calculation. The three-loop effective field theory that we are now studying is 
certainly more accurate, as it extends the neighborhood of the tricritical fixed point in which 
perturbation theory can be applied. In fact, the three-loop calculation serves several purposes.

One conclusion from the three-loop calculation is that it demonstrates the 
universality of the renormalization group at this order, that is to say, 
its independence from the cutoff scheme employed. 
Of course, the renormalization process is scheme-dependent, although 
suitable cutoff schemes do not result in a significant difference, 
as shown in our preceding article. 
In the three-loop calculation, we have further shown how to 
reduce non-universal renormalization terms to logarithms. 
In fact, we have found a single non-universal term that merges 
$\log\L$ with the scheme-dependent constant from the integral $I_3$, and this term 
vanishes in the beta functions. This vanishing is, of course, related to 
a general property of beta functions in ordinary renormalization theory: their finiteness,  
which forbids the presence of a $\log\L$ in them.

Reducing non-universal terms to logarithms is the simplest possible scheme. 
Furthermore, this scheme links the renormalization of the effective field theory 
to the most standard renormalization method,  
based on dimensional regularization.  
Previous work applying this method to sextic field theory yields incomplete results that 
cannot describe the tricritical-critical crossover. 
To do so, it is necessary to include in the renormalization process 
all terms affecting the two distinct fixed points. That is, 
finite terms must be included as well as
$1/\ve$ poles, the latter being equivalent to $\log\L$ terms in effective field theory. 
This is obvious once we consider that the renormalization of the quartic theory can be carried out without dealing with divergences 
($1/\ve$ poles or $\log\L$ terms) 
Finite terms have been systematically discarded  
in the sextic theory, 
thus preventing its use in describing the crossover in three dimensions. 
Furthermore, the entire mass range, from zero to infinity, must be considered. Therefore, 
an expansion in powers of mass is not valid. In fact, certain terms required in the 
renormalization formulas are non-analytic with respect to mass.

Despite the reduction of non-universal terms to logarithms, positive powers of the cutoff 
still play a role, even without contributing to beta functions. 
In fact, terms with positive powers of the cutoff can be obtained from these beta functions, in 
a generic form. The positive powers appear when the bare coupling constants are assumed to be 
essentially equivalent to the initial conditions for the renormalization group flow at the 
cutoff scale. 
This assumption becomes a condition for comparing perturbative and non-perturbative
formulations of the renormalization group. 

In any event, our main focus is on the nature of the renormalization group flow that we find.
We apply standard methods from dynamical systems theory to the analysis of this flow, 
in the plane of dimensionless coupling constants. 
Therefore, we base our conclusions on new beta functions and on new methods of analysis. 
Fortunately, planar dynamical systems are simple to study theoretically 
and simple to plot. We can easily determine that 
the tricritical fixed point (at the origin) is a saddle-node singular point. 
Our next step is to find the separatrices. 
The separatrix $g(u)$ tangent to the $u$-axis had been already studied in Ref.~\onlinecite{I}, 
as the flow line connecting the two fixed points. 
The three-loop beta functions allow us to derive the corresponding three-loop term of $g(u)$,
which is in full accord with the results of other methods of calculation.

The second separatrix primarily affects tricritical behaviour, since, 
in our regularization scheme, it is asociated with pure sextic theory (a potential with $\l_0=0$). 
Together with the first separatrix, it delimits the hyperbolic sector essential for the 
crossover. This second separatrix could also be a connecting line, 
if another non-trivial fixed point were to be found on it, as 
has been speculated in the literature (the evidence is negative). 
In any case, the flow along this separatrix, given by a restricted beta function, 
can be compared to that of the well-known beta function of pure sextic theory. 
Unfortunately, we cannot perform an effective 
comparison, as it requires the complete four-loop (or higher) order beta functions, 
which we have not calculated. While this calculation would be of interest for such a comparison,  
we do not expect it to deepen our understanding of the physics of the tricritical-critical crossover.

One more consequence of the theory of dynamical systems is that the {\em normal form}  
of the RG flow is relatively simple, as befits a saddle-node singularity. The first beta 
function can be transformed into a linear function, while the beta function of the marginal coupling 
can be transformed into a third-degree polynomial, both transformations being infinitely 
differentiable. Such normal form can be explicitly integrated in terms of known functions, thus 
fully revealing the nature of the singularity.

Some extensions or generalizations of this work may broaden its scope. It would be interesting 
to conduct a more in-depth comparison between perturbative and non-perturbative (exact) formulations 
of the renormalization group, to evaluate their performance in this simple field theory and to 
attempt to apply the conclusions to more complex field theories (other critical phenomena, the 
standard model of particle physics, gravity, etc.).

A different generalization of this work that may shed new light on scalar field theory is its extension to a larger set of bare coupling constants. 
While this extension takes us outside the scope of 
renormalizable three-dimensional scalar field theory, it is permissible within the framework of 
effective field theories. Indeed, we have argued that this extension could be useful for the study of 
broken symmetry phases. This will be addressed in future work.






\begin{acknowledgments}
I am grateful to Riccardo Guida and Andrey Kudlis for remarks on the renormalization 
of Phi$^4$ theory. 
\end{acknowledgments}

\appendix

\section{Calculation of integrals}
\label{integrals}

The two-loop integrals have been studied in various cutoff schemes in Ref.~\onlinecite{I}. 
Rajantie \cite{Raj} has calculated the 3-loop integrals of $(\lambda\phi^4)_3$ theory 
with dimensional regularization, 
allowing for different masses in the propagators. 
We could employ the preceding results, but it is useful to present what 
probably is the simplest cutoff calculation, in position space, taking advantage of 
the simple form of the propagator in position space (as does Rajantie \cite{Raj}):
$$
G_M(r) = \frac{\exp(-Mr)}{4\pi r}\,.
$$
(see also \cite[Appendix to Ch 5]{Parisi}). 

For example, 
$$
I_{2}(M) = 
\int\frac{d^{3}{k}}{\left(2\pi\right)^{3}\left({k}^{2}+M^{2}\right)}
= G_M(0) = \langle \phi^2\rangle,
$$
giving rise to the basic normal-order divergence $\langle \phi^2\rangle_\mathrm{div}$. Of course,
$$
G_m(r) = \frac{e^{-mr}}{4\pi r} = \frac{1}{4\pi r} - \frac{m}{4\pi} + O(m^2).
$$
Therefore, the regularized integral $I_{2}$ is:
$$
\langle \phi^2\rangle - \langle \phi^2\rangle_\mathrm{div} = G_M(0) - G_0(0) = 
- \frac{M}{4\pi} \left(1+ O\left(\frac{M}{\L}\right)\right).
$$

$I_1$ is more divergent than $I_2$, but its divergence can be reduced by taking its derivative 
with respect to $M$ \cite[App.\ A]{I}. Thus, we obtain:
$$
I_{1}(M) = \int I_{2}(M)\, 2M\,dM = - \frac{M^3}{6\pi} \left(1+ O\left(\frac{M}{\L}\right)\right).
$$

Let us proceed to the logarithmic divergences. The basic 
logarithmic divergent integral is $I_3$: 
\begin{equation*}
I_{3}\!\left(M\right) =  \int d^3\!x \left[G_M(x)\right]^3
= \frac{1}{16\pi^2} \int_0^\infty \frac{dr}{r} \exp(-3Mr).
\end{equation*}
The integral is divergent at $r=0$ (a UV divergence). With a cutoff $\L$, it yields: 
\begin{equation}
I_{3}\!\left(M\right) =
\frac{1}{16\pi^2}\left(\log\frac{\L}{M} - A + O\left(\frac{M}{\L}\right)\right).
\label{I3G}
\end{equation}
The constant $A$ is scheme-dependent but generally $1.5<A<2.5$ \cite{I}.
For example, we can use the simple space-cutoff scheme:
$$I_{3}\!\left(M\right) = \frac{1}{16\pi^2} \int_{1/\L}^\infty \frac{dr}{r} \exp(-3Mr).
$$
Now we have a standard integral \cite{GR}:
$$
\int_{1/\L}^\infty \frac{dr}{r} \exp(-3Mr) = \int_{3M/\L}^\infty \frac{dx}{x}\, e^{-x} 
= {E_1(3M/\L)} = -\gamma - \log 3 + \log\frac{\L}{M} + O\left(\frac{M}{\L}\right).
$$
Altough we have, in this scheme, $A = \gamma + \log 3 = 1.67583$, 
it is not necessary to specify it. In the main text, $A$ also includes the term $\log\L$. 

The other integral with a logarithmic divergence is $I_4$
(corresponding to another ``melon'' Feynman graph). To wit:
\begin{equation}
I_{4}\!\left(M\right) =  \int d^3\!x \left[G_M(x)\right]^4
= \frac{1}{64\pi^3} \int_0^\infty \frac{dr}{r^2} \exp(-4Mr).
\end{equation}
This integral is more divergent than $I_3$. Again, by taking 
its derivative with respect to $M$, its divergence is reduced to a logarithmic divergence. 
Actually, we obtain essentially the same integral $I_3$, which we replace by the 
regularized integral:
\begin{align}
\frac{dI_{4}(M)}{dM} &= \frac{-4}{64\pi^3} \int_0^\infty \frac{dr}{r} \exp(-4Mr) = 
-\frac{1}{16\pi^3} {E_1(4M/\L)}  \nonumber\\
 &= -\frac{1}{16\pi^3} \left[-\gamma - \log 4 + \log\frac{\L}{M} + O\left(\frac{M}{\L}\right)\right].
\end{align}
Hence,
\begin{equation}
I_{4}(M) = -\frac{1}{16\pi^3} \left[(-\gamma - \log 4)\,M + M\left(\log\frac{\L}{M} + 1 
+ O\left(\frac{M}{\L}\right)\right)\right].
\label{I4G}
\end{equation}
We have omitted the $M$-independent integration constant, which is proportional to $\L$. 

According to Eq.~(\ref{Vef}), $I_4$ is to be combined with $I_5$, which does not have 
logarithmic divergences and is:
$$
I_5 = I_2^2 \int d^3\!x \left[G_M(x)\right]^2 = \left(\frac{M}{4\pi}\right)^2
\frac{1}{4\pi} \int_0^\infty dr \exp(-2Mr) = \frac{M}{128\,\pi^3}\,.
$$
Therefore, the combination of integrals in Eq.~(\ref{Vef}) is:
$$
I_4 + 3 I_5 = \frac{1}{16\pi^3} \left[\left(\gamma + \log 4 - 1 + \frac{3}{8}\right)M 
- M\log\frac{\L}{M}\right] = \frac{M}{16\pi^3} \left(B + \log M\right).
$$
Here, we have defined
$$
B = \gamma + \log 4 - \frac{5}{8} - \log\L = A + \log \frac{4}{3} - \frac{5}{8}\,,
$$
where $A$ includes the term $\log\L$, as in the main text.

The next integral, $I_6$, contains a normal-order divergence and is factorized as 
$I_6=I_2\,I'_6$, where $I'_6$ corresponds to a 3-point two-loop Feynman graph. The 
(finite) graph integral $I'_6$ can be obtained from $I_3$, which we can write in Fourier space as:
$$
I_{3}\!\left(M\right)= \frac{1}{\left(2\pi\right)^{9}}
\int
\frac{d^3\!p\, d^3k\, d^3\!q\,\d(\bm{p}+\bm{k}+\bm{q})}
{\left({p}^{2}+M^{2}\right)\left({k}^{2}+M^{2}\right)\left({q}^{2}+M^{2}\right)}\,,
$$
We have:
$$
I'_6 = 
\frac{1}{\left(2\pi\right)^{9}}
\int \frac{d^3\!p\, d^3k\, d^3\!q\,\d(\bm{p}+\bm{k}+\bm{q})}
{\left({p}^{2}+M^{2}\right)^2\left({k}^{2}+M^{2}\right)\left({q}^{2}+M^{2}\right)}
= -\frac{1}{3}\,\frac{dI_{3}}{dM^2} = \frac{1}{\left(4\pi\right)^2}\, \frac{1}{6 M^2}\,,
$$
where we have used Eq.~(\ref{I3G}). Therefore,
$$
I_6=I_2\,I'_6 = -\frac{M}{4\pi}\,\frac{1}{\left(4\pi\right)^2}\, \frac{1}{6 M^2} 
= \frac{-1}{\left(4\pi\right)^3}\, \frac{1}{6 M}\,.
$$

The remaining integrals, namely, $I_7$, $I_8$, and $I_9$, are finite but difficult to 
calculate. We take them from Ref.~\onlinecite{Raj}. 
The first one, $I_7$, is to be added to $I_6$. Let us quote:
\begin{align*}
I_7 &= \frac{1}{\left(4\pi\right)^3} 
   \left(\frac{\pi ^2}{12}+\text{Li}_2\left(-\frac{1}{3}\right)\right)/M,\\
I_8 &= \frac{1}{\left(4\pi\right)^3} \left[\frac{1}{2}\text{Li}_2\left(-\frac{1}{3}\right)
  +\frac{\pi^2}{24}-\frac{2}{3} \log \left(\frac{4}{3}\right)\right]/M^{3}.
\end{align*}
The last one, $I_9$, is computed in terms of the dilogarithm function and a definite integral.  
The complete result can be expressed in terms of the following numbers (using the notation of 
Ref.~\onlinecite{Guida-ZJ}):
\begin{align}
k_3 &=-\frac{9}{2}+\frac{9 \pi ^2}{4}-\frac{27}{2} \log ^2\left(\frac{4}{3}\right)
-27 \text{Li}_2\left(\frac{1}{4}\right)=
9.36272,\\
k_4 &=
\frac{81 \text{Li}_2\left(\frac{1}{4}\right)}{2}-\frac{27 \pi ^2}{8}+\frac{81}{4} \log ^2\left(\frac{4}{3}\right)+54 \log \left(\frac{4}{3}\right) \nonumber\\
&\phantom{==}-27 \sqrt{2} 
\int_0^1 \frac{dx}{\sqrt{3-x^2}}
\left({-\frac{x^2 \log \left(\frac{4}{x+2}\right)}{4-x^2}+\frac{x \log \left(\frac{x+3}{3}\right)}{x+2}+\log \left(\frac{x+3}{x+2}\right)-\log
   \left(\frac{4}{3}\right)}\right) \nonumber\\
 &= -6.43307.
\end{align}
Let us note that $k_3$ and $k_4$ are proportional, respectively, to the combinations $I_6+I_7$ 
and $3I_8+2I_9$ in Eq.~(\ref{Vef}).

\subsection{Integrals for $Z$}
\label{I4Z}

We still need some further integrals for the field renormalization factor $Z$.
Since the integrals for the calculation of $Z$ in the quartic theory 
are finite and are given by Kudlis and Pikelner \cite{Kudlis}, we focus on 
the tenth Feynman graph integral, which is the only remaining one, for 
the sextic theory. It must have two (amputated) external legs with  
an external wavenumber $p$. 
Evidently, the integral factorizes and we can express it in terms of  
the third graph integral, which we need with an external wavenumber $p$. 

Therefore, we must calculate
\begin{equation*}
I_{3}\!\left(M,p\right) =  \int d^3\!x \exp(-i\,\bm{p}\cdot\bm{x})\left[G_M(x)\right]^3 
= \frac{1}{16\pi^2} \int_0^\infty \frac{dr}{r^2}\, \frac{\sin(p r)}{p} \exp(-3Mr).
\end{equation*}
Let us expand the integrand in a power series:
$$
\int_0^\infty \frac{dr}{r^2}\, \frac{\sin(pr)}{p} \exp(-3Mr) = 
\int_0^\infty \frac{dr}{r}\,\exp(-3Mr) - 
\frac{p^2}{6} \int_0^\infty {dr}\,r  \exp(-3Mr) + O(p^4).
$$
The first integral is divergent and has been calculated for Eq.~(\ref{I3G}). The second 
integral is finite:
$$
\int_0^\infty {dr}\,r \exp(-3Mr) = \frac{1}{9 M^2}\,.
$$
In total:
\begin{equation*}
I_{3}\!\left(M,p\right) 
= \frac{1}{16\pi^2} \left(\log\frac{\L}{M} - A - \frac{p^2}{54 M^2}\right).
\end{equation*}

Of course, the above integral is to be multiplied by $I_2$ and
contributes with the coupling constants and combinatorial factor of the corresponding term 
in Eq.~(\ref{Vef}). To wit:
\begin{align}
-\frac{U^{(5)}}{5!} \frac{U^{(3)}}{3!}\,{60}\, I_2\,I_{3}\!\left(M,p\right) = 
-\frac{U^{(5)}U^{(3)}}{12} \left(-\frac{M}{4\pi}\right) \frac{1}{16\pi^2}
\left(\log\frac{\L}{M} - A - \frac{p^2}{54 M^2}\right). 
\end{align}
The $p$-independent part matches the coresponding term in Eq.~(\ref{Veff}). 

We have, 
besides the $p$-independent term, a $p^2$ term that gives 
$$\d Z^{-1} =  -\frac{5\,\l_0\,g_0}{6\pi^3\,{m_0}}\,.
$$
Note that we replace $m_0$ with $\widetilde{m}$ when we insert $\d Z^{-1}$ in the complete expression 
of $Z^{-1}$ in the main text. This is allowed in the present perturbation order.

\section{Full renormalization formulas}
\label{fren}

The renormalization equations directly obtained from the effective potential 
have been much abbreviated in the main text, as Eqs.~(\ref{mtm0}), (\ref{ll0}) and (\ref{gg0}). 
Here we display them in full:
\begin{align}
Z^{-1} m^2 &= U_\mathrm{eff}''(0) =
m_0^2-\frac{3 h \left(m_0 \lambda _0\right)}{\pi } + \frac{h^2
   \left(12 (4 A+3) \lambda _0^2+45 g_0 m_0^2+48 \lambda _0^2 \log m_0\right)}{8 \pi ^2}
   \nonumber\\
   &- \frac{h^3\lambda _0}{16 \pi ^3 m_0}\left\{
   45 g_0 m_0^2\, (16 A+16 B+3)+16 \lambda _0^2\, (9 B-2
   {k_3}+9)\right.
\nonumber\\
&\left.
   {}+ 144 \log m_0 \left(10 g_0 m_0^2+\lambda _0^2\right)
   \right\}, 
\label{mtm0A}\\
Z^{-2} \lambda &= \frac{U_\mathrm{eff}^{(4)}(0)}{4!} = 
   \lambda _0-\frac{3 h \left(5 g_0 m_0^2+6 \lambda _0^2\right)}{4 \pi  m_0}
   +\frac{3 h^2 \lambda _0 }{8 \pi ^2 }
   \left(5 (16 A+21) g_0 +80 g_0 \log m_0+\frac{48\lambda _0^2}{m_0^2} \right)
   \nonumber\\
   &- \frac{h^3}{64 \pi ^3 m_0^3}\left\{
   10 g_0 m_0^2 \lambda _0^2\, (864 A+936 B-160 {k_3}+1881)+225
   g_0^2 m_0^4\, (32 A+48 B+3)\right.
\nonumber\\
&\left.
   {}-32 \lambda _0^4\, (27 B-12 {k_3}+8 {k_4})+144 \log m_0 \left(125
   g_0 m_0^2 \lambda _0^2+125 g_0^2 m_0^4-6 \lambda _0^4\right)
   \right\},
\label{ll0A}\\
Z^{-3} g &= \frac{U_\mathrm{eff}^{(6)}(0)}{6!} = g_0 + \frac{9 h \lambda _0 
\left(2 \lambda _0^2-5 g_0 m_0^2\right)}{2 \pi  m_0^3} + \frac{h^2}{8 \pi ^2 } 
\left( 75 (8 A+9) g_0^2+600 g_0^2
   \log m_0+\frac{1800 g_0 \lambda _0^2}{m_0^2}\right.
\nonumber\\
&\left.
   {}-\frac{864 \lambda _0^4}{m_0^4} \right) -\frac{h^3 \lambda _0}{32 \pi ^3 m_0^5}
   \left\{-10 g_0 m_0^2 \lambda _0^2\, (1296 A+1512 B-528 {k_3}+256
   {k_4}+81)\right.
\nonumber\\
&\left.
   {}+ 25 g_0^2 m_0^4\, (1296 A+1728 B-224
   {k_3}+3105)+288 \lambda _0^4\, (9 B-6 {k_3}+8
   {k_4}-3)
   \right.
\nonumber\\
&\left.
   {}+432 \log m_0 \left(-65 g_0 m_0^2
   \lambda _0^2+175 g_0^2 m_0^4+6 \lambda _0^4\right)
   \right\}.
\label{gg0A}
\end{align}
[The symbol $O\!\left(h^4\right)$ is omitted, for brevity.]

\section{Beta functions} 
\label{betas}

To calculate the beta-functions, we proceed as in Ref.~\onlinecite{I}. 
We must first obtain the renormalized coupling constants in terms of the renormalized mass and 
bare coupling constants, by combining 
equations (\ref{ltl0}), (\ref{gtg0}), (\ref{Zi}), and (\ref{mtm+}). 
The result is:
\begin{align}
\l &= 
\lambda _0 - \frac{h}{4 \pi}\left(15 g_0 {m}+\frac{18 \lambda _0^2}{{m}}\right) +
\frac{h^2}{16 \pi^2}\left(
60 g_0 \lambda _0 (8 A+8 \log m + 9)+\frac{3532 \lambda _0^3}{9 m^2}   
\right)
\nonumber\\
&\phantom{==} +{}
\frac{h^3}{64 \pi^3}\left\{
-\frac{20 g_0 \lambda _0^2 }{3 m}
\left( 2592 A-240 {k_3}+1571+1404 \log \left(\frac{4}{3}\right)+2592 \log m\right) -
\right.
\nonumber\\
&
\left.   
{450 g_0^2 m}
    \left(40 A-15+24 \log \left(\frac{4}{3}\right)+ 40 \log m\right)
   -\frac{8 \lambda _0^4}{{m}^3}
   \left(48 {k_3}-32 {k_4}-8 {k_5}+611-108 \log \left(\frac{4}{3}\right)
   \right)
\right\} \nonumber\\
&\phantom{==} +{} O\!\left(h^4\right),
\label{lrl0}
\end{align}
\begin{align}
g = g_0 &+ \frac{h}{4 \pi}\left(\frac{36 \lambda _0^3}{{m}^3}-\frac{90 g_0 \lambda _0}{{m}}\right) 
 +\frac{h^2}{16 \pi^2}\left(150 g_0^2 \left(8 \log {m}+8 A+9\right) +
  \frac{12404 g_0 \lambda _0^2}{3 m^2}-\frac{2376
   \lambda _0^4}{m^4}   \right)
\nonumber\\
&+ 
\frac{h^3}{64 \pi^3}\left\{
\frac{16 g_0 \lambda _0^3}{{m}^3} \left(
3240 A-660 {k_3}+320 {k_4}+6 {k_5}+3240 \log m-3745+1890 \log \left(\frac{4}{3}\right)
   \right)
\right.
\nonumber\\
&
\left. \phantom{===}
   {}- \frac{20 g_0^2 \lambda _0}{{m}}\left(
7560 A-560 {k_3}+7560 \log m + 4897+4320 \log
   \left(\frac{4}{3}\right)   \right)
\right.
\nonumber\\
&
\left.   \phantom{===}
   {}+ \frac{576 \lambda _0^5}{{m}^5} \left(
   6 {k_3}-8 {k_4}+97-9 \log \left(\frac{4}{3}\right)
   \right)
   \right\}  + O\!\left(h^4\right),
\label{grg0}
\end{align}

Now we can take the derivatives of the renormalized coupling constants with respect to $m$, 
keeping fixed $\l_0$ and $g_0$. This operation is straightforward, but the resulting derivatives 
are expressed as functions of the bare coupling constants. To express them as functions of 
the renormalized coupling constants, we need to invert 
Equations (\ref{lrl0}) and (\ref{grg0}) up to $O\!\left(h^2\right)$ and substitute for 
$\l_0$ and $g_0$.
Thus, we obtain:
\begin{align}
\left(\frac{\p \l}{\p m}\right)_{\!\!\l_0,\,g_0} &=
h \left(\frac{9 \lambda^2}{2 \pi  m^2}-\frac{15 g}{4 \pi
   }\right)+
h^2\left(\frac{907 \lambda ^3}{36 \pi ^2 m^3}-\frac{165 g \lambda }{8 \pi
   ^2 m}\right) +
\nonumber\\
&h^3\left(
-\frac{675 g^2 \log \left(\frac{4}{3}\right)}{4 \pi ^3}+\frac{g
   \lambda ^2 \left(-1200 {k_3}+6415+7020 \log
   \left(\frac{4}{3}\right)\right)}{48 \pi ^3 m^2}
   \right.
\nonumber\\
&
\left. \phantom{==}
{}-\frac{\lambda
   ^4 \left(-144 {k_3}+96 {k_4}+24 {k_5}+911+324
   \log \left(\frac{4}{3}\right)\right)}{8 \pi ^3 m^4}
   \right)   
   +O\left(h^4\right),
\\
\left(\frac{\p g}{\p m}\right)_{\!\!\l_0,\,g_0} &=
h \left(\frac{45 g \lambda }{2 \pi  m^2}-\frac{27 \lambda^3}{\pi 
   m^4}\right)
+ h^2\left(
\frac{1275 g^2}{8 \pi ^2 m}-\frac{2557 g \lambda ^2}{12 \pi ^2
   m^3}+\frac{27 \lambda ^4}{\pi ^2 m^5}
\right) 
\nonumber\\
&h^3\left(
\frac{5 g^2 \lambda  \left(-280 {k_3}+2119+2160 \log
   \left(\frac{4}{3}\right)\right)}{8 \pi ^3 m^2} +{}
   \right.
\nonumber\\
&
\left. \phantom{===}   
   \frac{3
   \lambda ^5 \left(-360 {k_3}+480 {k_4}+7891+540 \log
   \left(\frac{4}{3}\right)\right)}{4 \pi ^3 m^6} -{}
   \right.
\nonumber\\
&
\left. \phantom{==}
\frac{g \lambda ^3 \left(-3960 {k_3}+1920 {k_4}+36
   {k_5}+63797+11340 \log \left(\frac{4}{3}\right)\right)}{8
   \pi ^3 m^4}
   \right)   
   +O\left(h^4\right),
\end{align}
Let us notice that the scheme dependence of Eqs~(\ref{lrl0}) and (\ref{grg0}), due to 
the constant $A$, has disappeared.

The beta functions~(\ref{b1}) and (\ref{b2}) of dimensionless coupling constants 
$u=\l/m$ and $g$ are easily obtained by substitution and taking into account:
\begin{align*}
\b_1 &= m\left(\frac{\p u}{\p m}\right)_{\!\!\l_0,\,g_0} =
\left(\frac{\p \l}{\p m}\right)_{\!\!\l_0,\,g_0}\!\! - \frac{\l}{m} \,.
\end{align*}

\end{document}